\documentclass[12pt,document,nofootinbib,superscriptaddress,onecolumn,preprintnumbers,balancelastpage]{revtex4}
\pdfoutput=1
\usepackage{latexsym}
\usepackage{amssymb}
\usepackage{epsfig,amsmath,graphics}
\usepackage{listings}
\usepackage{color}
\usepackage{dcolumn}
\usepackage{slashed}
\usepackage{cancel}
\usepackage{comment,latexsym} 
\usepackage{bm}
\usepackage{verbatim}
\usepackage{tabularx}
\usepackage{dcolumn}
\definecolor{Mahogany}{rgb}{0.62,0.24,0.15}
\definecolor{colorLink}{rgb}{0.7,0,0}
\definecolor{colorCite}{rgb}{0,.7,0}
\definecolor{colorURL}{rgb}{0,0,0.7}
\definecolor{colorTC}{rgb}{.2,.7,.2}
\definecolor{colorDP}{rgb}{.7,.7,.2}
\usepackage[colorlinks=true,linktocpage=true,linkcolor=black,citecolor=colorCite,urlcolor=colorURL]{hyperref}
\usepackage{setspace}
\usepackage{xspace}
\usepackage{multirow}
\usepackage{titlesec}
\usepackage[bbgreekl]{mathbbol}
\usepackage{bm}
\usepackage{float}
\usepackage{placeins}
\usepackage{afterpage}

\graphicspath{{Figures/}}

\newcommand{\Lc}{\mathcal{L}}

\def\be{\begin{equation}}
\def\ee{\end{equation}}
\newcommand{\beq}{\begin{equation}}
\newcommand{\eeq}{\end{equation}}
\newcommand{\eref}[1]{Eq.~(\ref{#1})}

\newcommand{\lsim}{\!\mathrel{\hbox{\rlap{\lower.55ex \hbox{$\sim$}} \kern-.34em \raise.4ex \hbox{$<$}}}}
\newcommand{\gsim}{\!\mathrel{\hbox{\rlap{\lower.55ex \hbox{$\sim$}} \kern-.34em \raise.4ex \hbox{$>$}}}}
\newcommand{\vev}[1]{ \left\langle {#1} \right\rangle }
\newcommand{\GeV}{{\text{ GeV}}}

\newcommand{\gev}{{\text{ GeV}}}
\newcommand{\tev}{{\text{ TeV}}}

\newcommand{\llog}[1]{\log \left[ \frac{\widetilde{m}^2_{#1} }{ v^2}\right]}
\newcommand{\m}[1]{\widetilde{m}^2_{#1}}

\expandafter\def\expandafter\normalsize\expandafter{%
    \normalsize
    \setlength\abovedisplayskip{8pt}
    \setlength\belowdisplayskip{8pt}
    \setlength\abovedisplayshortskip{8pt}
    \setlength\belowdisplayshortskip{8pt}
}

\titleformat{\section}{\center\normalfont\fontsize{14}{15}\bfseries}{\thesection.}{1em}{}
\titleformat{\subsubsection}{\center\normalfont\fontsize{12}{15}}{\thesubsubsection.}{1em}{}

\begin{document}
$\quad$
 \begin{flushright} 
NSF-KITP-16-058
  \end{flushright} 
\vskip 60 pt

\title{750 GeV Diphotons from Supersymmetry with Dirac Gauginos} 

\author{Timothy Cohen}
\affiliation{
Institute of Theoretical Science, University of Oregon, Eugene, Oregon 97403
\vspace{-4pt}
}

\author{Graham D. Kribs}
\affiliation{
Institute of Theoretical Science, University of Oregon, Eugene, Oregon 97403
\vspace{-4pt}
}

\author{Ann E. Nelson}
\affiliation{
Department of Physics, University of Washington, Seattle, Washington 98195
\vspace{-4pt}
}

\author{Bryan Ostdiek$^{\,}$}
\affiliation{
Institute of Theoretical Science, University of Oregon, Eugene, Oregon 97403
\vspace{-4pt}
}

\begin{abstract}
\vskip 40 pt
\begin{center}
{\bf Abstract}
\end{center}
\vskip -30 pt
$\quad$
\begin{spacing}{1.05}\noindent
Motivated by the recent excess in the diphoton invariant mass near 
750 GeV, we explore a supersymmetric extension of the Standard Model 
that includes the minimal set of superpartners as well as additional Dirac partner chiral superfields in 
the adjoint representation for each gauge group.  The bino partner
pseudoscalar is identified as the 750 GeV resonance, while 
superpotential interactions between it and the gluino (wino) 
partners yield production via gluon fusion (decay to photon pairs) 
at one-loop.  The gauginos and these additional adjoint superpartners are married by a Dirac mass and must also
  have Majorana masses.  While a large wino partner Majorana mass is necessary to explain the excess, the gluino can be approximately Dirac-like, 
providing benefits consistent with being both ``supersoft'' 
(loop corrections to the scalar masses from Dirac gauginos 
are free of logarithmic enhancements) and ``supersafe'' 
(the experimental limits on the squark/gluino masses can be 
relaxed due to the reduced production rate).
Consistency with the measured Standard Model-like Higgs boson mass 
is imposed, and a numerical exploration of the parameter space 
is provided.  Models that can account for the diphoton excess are additionally characterized by having couplings that
can remain perturbative up to very high scales,
while remaining consistent with experimental constraints, the Higgs boson mass, and an electroweak scale which is not excessively fine-tuned.   
\end{spacing}
\end{abstract}

\maketitle
\newpage
\begin{spacing}{1.3}
\pagebreak
%
%

\section{Introduction}
\label{sec:Intro}

LHC run II is upon us.  The first round of results has already yielded a tantalizing hint of new physics, which has first appeared as an excess in the diphoton invariant mass distribution at $\sim 750 \GeV$.  In this paper, we explore the possibility that this hint could be the first manifestation of a larger supersymmetric structure that yields a natural electroweak scale while accommodating a Standard Model--like Higgs boson at 125 GeV.  Specifically, the focus here will be on minimal models with Dirac gauginos --- the requisite Dirac partner states will provide the ingredients necessary to explain the diphoton excess.

The largest excess is due to the 13 TeV ATLAS data set (3.9$\sigma$ local)~\cite{ATLAS-CONF-2015-081, ATLAS-CONF-2016-018}, which is only in mild tension with the lack of an observation by ATLAS at 8 TeV~\cite{Aad:2014ioa,Aad:2015mna}.  There is a smaller complimentary excess in the 13 TeV CMS data (2.9$\sigma$ local)~\cite{CMS-PAS-EXO-15-004, CMS-PAS-EXO-16-018}, which grows to 3.4$\sigma$ local when combined with their 8 TeV data~\cite{CMS-PAS-EXO-16-018, Khachatryan:2015qba, Chakrabortty:2015hff}.  A statistical combination of all these results (performed external to either collaboration) yields 4$\sigma$ local, with a best fit cross section of $\sigma \times \text{BR}_{\gamma\gamma} \simeq 3.8 \text{ fb}$ at 750 GeV~\cite{Buckley:2016mbr, BuckleyBlog}, see also~\cite{Franceschini:2015kwy, Knapen:2015dap, Agrawal:2015dbf, Gupta:2015zzs, Altmannshofer:2015xfo, Craig:2015lra, Bernon:2016dow, Kamenik:2016tuv, Franceschini:2016gxv} which discuss  interpretations of the data.  While there is a hint that the excess is quite wide, we will assume that the narrow width approximation is appropriate for the rest of this paper.  Fortunately, more data is on the way, and so the experimental status will evolve dramatically over the next few years.  In anticipation of this update, it is interesting to explore  concrete scenarios that allow us to investigate how plausible such a signal is in a given model.

There have been a staggering number of papers probing this question,\footnote{According to~\cite{Backovic:2016xno}, more are expected to be written.} see~\cite{Staub:2016dxq} for a summary of many of the proposed models in the context of a toolkit to explore their predictions.  Many authors have chosen to take a  phenomenologically motivated approach by introducing a variety of states whose sole purpose is to provide the 750 GeV state itself, a coupling to either quarks or gluons, and a coupling to photons.  In the simplest scenarios, the excess results from a scalar or pseudoscalar decaying to a pair of photons.  This new particle cannot carry electric charge implying that its coupling to photons must be due to a higher dimensional operator.  One simple mechanism is to induce this operator via loops involving additional beyond the Standard Model states.  The model building becomes less trivial, and one or more of the following is required: large couplings, many states running in the loop, and/or large electric charges for these new states.  Theories of these types do not have to result from a UV biased approach to model building, although  there is a growing literature of models which are concerned with incorporating the new states in the context of supersymmetry (SUSY)~\cite{Petersson:2015mkr, Demidov:2015zqn, Carpenter:2015ucu, Feng:2015wil, Ding:2015rxx, Wang:2015kuj, Chakraborty:2015gyj, Allanach:2015ixl, Casas:2015blx, Hall:2015xds, Hall:2016swn, Wang:2015omi, Tang:2015eko, Chao:2016mtn, Dutta:2016jqn, King:2016wep, Ding:2016udc, Ellwanger:2016qax, Han:2016fli, Barbieri:2016cnt, Badziak:2016cfd, Baratella:2016daa, Nilles:2016bjl}, and consistency of the models to high energies by studying their renormalization group evolution~\cite{Gu:2015lxj, Hall:2015xds, Bae:2016xni, Hamada:2016vwk, Barbieri:2016cnt, Han:2016fli, Choudhury:2016jbc, Djouadi:2016oey, Hall:2016swn, Bardhan:2016rsb,Bharucha:2016jyr}.

In this paper, we will explore the connection between  the diphoton excess and  the  extension of the  Minimal Supersymmetric Standard Model (MSSM) with Dirac gauginos~\cite{Fox:2002bu}.  This class of supersymmetric models continues to survive  experimental constraints while achieving a natural electroweak scale, for some recent developments see~\cite{Itoyama:2011zi, Itoyama:2013sn, Itoyama:2013vxa, Alves:2015kia, Martin:2015eca}.  This persistence is due to two effects known as supersoft~\cite{Fox:2002bu} and supersafe~\cite{Kribs:2012gx}.  The first refers to a property of the renormalization group equations (RGEs) for Dirac gaugino masses which can be understood as resulting from a certain class of spurions.  In the absence of Majorana masses for the gauginos, the gauge corrections to the scalar soft-mass spectrum are all finite. Thus there are no logarithmically enhanced corrections to scalar masses, which allows the gaugino masses to be rather heavy without fine-tuning.  The second feature of these models is that heavier Dirac gluinos imply a substantially suppressed production cross section of squarks at the LHC due to the absence of $t$-channel diagrams with a Majorana mass insertion on the internal gluino line.  Therefore  LHC bounds on superpartners in these models still leave a  natural parameter space open for experimental exploration.  For a study of how these conclusions change in the presence of both Dirac and Majorana masses for the gluinos, which is relevant for the parameter space explored below, see~\cite{Kribs:2013eua}.  Given these features and their relation to electroweak naturalness, it is interesting to understand if models with Dirac gauginos  could explain the diphoton excess.

Constructing the model requires extending the matter content of the MSSM to include an additional adjoint chiral superfield for each gauge group, along with a SUSY breaking Dirac mass term that marries these new adjoints to the gauginos.  The chiral adjoints   provide  candidates for the new resonance, along with additional states to run in the loop and generate the higher dimension couplings to gluons and photons.  Specifically, the (pseudo)scalar of the bino partner is a  candidate 750 GeV state, while a superpotential coupling between it and the $SU(3)_c$ octet [$SU(2)_L$ triplet] yields the loop induced coupling to gluons [photons].  Explaining the excess will motivate relatively large values for these couplings.  As explained below, these operators require that the states be split away from the exact Dirac limit.  

In the spirit of including all possible allowed operators that could be relevant to the diphoton phenomenology, we also will include a coupling between the bino partner and the Higgs superfields, which is analogous to the Next-to-Minimal Supersymmetric Standard Model (NMSSM) superpotential, along with one involving the wino partner and the Higgses.  These will both lead to additional contributions to the Higgs quartic; compatibility with a 125 GeV Higgs boson will yield important constraints on the parameter space.  We will also verify that the model is under perturbative control, and will emphasize regions of parameter space that can approach the Planck scale before becoming strongly coupled.  It is worth noting that the gauge couplings do not unify without additional matter.  Addressing this is beyond the scope of this work, and should decouple from the phenomenology of interest here beyond the potential impact on the scale of the Landau poles.  Combining the diphoton excess, the Higgs mass, naturalness and experimental constraints will point to a region of parameter space that predicts additional superpartners, some of which  could be accessible at the LHC, and   all are  within reach of proposed future colliders~\cite{Cohen:2013xda, Cohen:2014hxa, Gori:2014oua, Arkani-Hamed:2015vfh, Bramante:2015una, Bramante:2014tba}.  If the diphoton excess is the first sign of beyond the Standard Model physics, then a rich program involving the discovery of many additional states should follow.

The rest of this paper is organized as follows.  In the next section we introduce the details of the model and provide a qualitative discussion of the relevant processes.  Then Sec.~\ref{sec:NumericalResults} gives the detailed numerical analysis of the parameter space.  Conclusions are then given in Sec.~\ref{sec:Conclusions}, and some technical details are given in the Appendix.

\section{The Model}
\label{sec:TheModel}

We  introduce the details of the model.  The matter content includes that of the MSSM as well as an adjoint partner superfield for each of the gauge multiplets: $S$, a singlet of $U(1)_Y$; $T$, a triplet of $SU(2)_L$; and $O$, an octet of $SU(3)_c$. We do not include the additional inert Higgsinos needed for a $U(1)_R$-symmetric model \cite{Kribs:2007ac}.  We write the component fields as:
\begin{eqnarray}
S = \left( \begin{array}{c} \psi_S \\ S_B + i\, A_B \end{array} \right) \quad 
T = \left( \begin{array}{c} \psi_T \\ S_T + i\, A_T \end{array} \right) \quad
O = \left( \begin{array}{c} \psi_O \\ S_O + i\, A_O \end{array} \right)
\end{eqnarray}
The pseudoscalar $A_B$ is identified as the 750 GeV resonance.\footnote{We use $A_B$ to denote the pseudoscalar of the $S$ chiral superfield to remind the reader that $S$ is the partner superfield of the bino, as opposed to the singlet of the NMSSM.}  The octet fermions will generate a one-loop coupling between $A_B$ and the gluons, while the charged Higgsinos and charged triplet fermions will generate a one-loop coupling between $A_B$ and the photons.

The full Lagrangian is
\begin{equation}
\mathcal{L} \; = \; \mathcal{L}_\text{kin} + \mathcal{L}_\text{Dirac} + \mathcal{L}_\text{Maj} + \left(\int \text{d}^2 \theta \,W + \text{h.c.}\right) + \mathcal{L}_\text{soft},
\end{equation}
where $ \mathcal{L}_\text{kin}$ are the kinetic terms for all the fields in the model (including the gauge interactions via covariant derivatives), $ \mathcal{L}_\text{Dirac}$ and $ \mathcal{L}_\text{Maj}$ are the Dirac and Majorana gaugino masses respectively, $W$ is the superpotential containing all allowed renormalizable couplings, and $\mathcal{L}_\text{soft}$ contains all the scalar soft masses, $A$- and $B$-terms (for the sake of brevity, we will not explicitly write down this part of the Lagrangian).  In detail, the Dirac masses for the gauginos are contained in 
\begin{equation}
\mathcal{L}_\text{Dirac} \; \supset \; \int d^2 \theta \left(\frac{1}{\Lambda_S} W_\alpha'\,W_B^\alpha\, S + \frac{1}{\Lambda_T} W_\alpha'\,W_W^\alpha\, T+\frac{1}{\Lambda_O} W_\alpha'\,W_G^\alpha\, O\right) + \text{h.c.},
\end{equation}
where $\vev{W'_\alpha} = \theta_\alpha\,D'$ is the spurion that yields the Dirac mass, and the scales $\Lambda_a$ are the UV scale where these terms are generated.  Then we will denote the Dirac mass parameters as
\begin{equation}
M^D_{S} \equiv \frac{1}{\Lambda_S} D', \quad M^D_{T} \equiv \frac{1}{\Lambda_T} D', \quad M^D_{O} \equiv \frac{1}{\Lambda_O} D'\,.
\end{equation}
We also include Majorana mass terms for the gauginos given by
\begin{equation}
\mathcal{L}_\text{Maj} \supset \int d^2 \theta \bigg(\frac{X}{\Lambda_1} W_{B,\alpha}W_B^\alpha + \frac{X}{\Lambda_2} W_{W,\alpha}W_W^\alpha+\frac{X}{\Lambda_3} W_{G,\alpha}W_3^\alpha \bigg)+\text{h.c.},
\end{equation}
where $\vev{X} = F\, \theta^2$ is another SUSY breaking spurion, and the $\Lambda_i$ are the UV scale where these terms are generated.  We will use the standard notation for the Majorana gaugino masses, $M_i$ with $i = 1,2,3$.  

This is the supersoft model generalized to include Majorana masses.  In addition, we include the most general renormalizable superpotential interactions:
\begin{eqnarray}
W & \supset &{} \zeta_S\,S + \frac{M_S}{2}\, S\,S + \frac{M_T}{2}\, T^a\,T^a+ \frac{M_O}{2} \,O^b\,O^b \notag\\
& &{} + \frac{\lambda_{SOO}}{2} \,S\, O^b\,O^b + \lambda_{SHH}\, S\, H_{u}\, H_{d} + \frac{\lambda_{STT}}{2} \,S\, T^a\,T^a\notag \\
& &{} +  \lambda_{THH} H_u\, T\, H_{d}+ \frac{\kappa}{6} S^3 + \frac{\kappa_O}{3} \,\text{Tr}(O\,O\,O) + W_\text{MSSM} \, , 
\label{eq:Superpotential}
\end{eqnarray}
where $a=1,2,3$ for $SU(2)_L$, $b=1 \ldots 8$ for $SU(3)_c$,  $\zeta_S$ is the tadpole for $S$, $M_S$, $M_T$, and $M_O$ are the superpotential masses for the adjoint chiral superfields, $\lambda_{SHH}$, $\lambda_{STT}$, $\lambda_{SOO}$, and $\lambda_{THH}$, $\kappa_S$, and $\kappa_O$ are all possible gauge-invariant Yukawa couplings, and $W_\text{MSSM}$ is the MSSM superpotential that includes the $\mu$-term and the Standard Model Yukawa couplings.  The three Yukawa  terms in the middle line are crucial for the diphoton phenomenology.  The parameter $\lambda_{SOO}$ will determine the coupling between $A_B$ and the gluons, while $\lambda_{STT}$ and $\lambda_{SHH}$ will determine the coupling between the $A_B$ and the photons.  For simplicity, in the following we will set $\zeta_S$, $\kappa_S$, and $\kappa_O$ to zero, since they play no role in the physics of interest.  It is in principle viable to have $\vev{S} \equiv v_S$ and $\vev{T} \equiv v_T$ be nonzero; in the following we will also assume these are negligible unless they are specifically mentioned. Note that the supersoft limit is when $M^D_{i} \gg M_j$ for each set $(i,\{j\})) = (S, \{S,1\}), (T, \{T,2\}), (O, \{O,3\})$.  

\subsection{Explaining the 750 GeV diphoton excess}
\label{sec:Explain}

There is an obvious (pseudo)scalar in this model that is a candidate resonance to explain the excess, namely the scalar component of the bino partner $S \ni S_B + i\,A_B$.  In principle, $S_B$ could explain the excess due to its trilinear couplings with scalar superpartners, see~\cite{Carpenter:2015ucu}.  However, the natural assumption is that there would be nontrivial mixing between $S_B$ and the Higgs, especially when the couplings are taken large enough to achieve the best fit cross section, leading to stringent constraints, both from requiring that $S_B \rightarrow t\,\bar{t}$ does not dominate and also that the measured Higgs properties are within errors.  Therefore,  we choose to focus our attention on $A_B$ as the candidate 750 GeV state and assume that $S_B$ is heavier.  

The leading coupling between $A_B$ and the gauge bosons can be derived from general arguments.  From the interactions of the neutral pion in QCD, 
as well as the interactions of the CP-odd Higgs boson in
supersymmetry \cite{Djouadi:2005gi}, it is well known that 
that integrating out heavy fermions of mass $M_f$ that interact with
a pseudoscalar yield effective interactions 
with gauge bosons at dimension-5:
\begin{eqnarray}
\mathcal{L}_\text{eff} \supset \sum_f \left[
\frac{\alpha\, C_B(f)}{8\, \pi\, c_W^2\, M_f} A_B\, F_{\mu\nu}\, \widetilde{F}^{\mu\nu}
+ \frac{\alpha \,C_W(f)}{8\, \pi\, s_W^2 \,M_f}\, A_B\, W_{\mu\nu}^a \widetilde{W}^{a \mu\nu}
+ \frac{\alpha_3 \,C_G(f)}{8\, \pi\, M_f} \,A_B\, G_{\mu\nu}^b \widetilde{G}^{b \mu\nu} 
\right] , 
\label{eq:anomalyterms}
\end{eqnarray}
where $F_{\mu\nu}$ is the $U(1)_Y$ field strength, $W_{\mu\nu}$ is the $SU(2)_L$ field strength, $G_{\mu\nu}$ is the $SU(3)_c$ field strength, $a=1,2,3$, $b=1 \ldots 8$, $s_W$ ($c_W$) are the sine (cosine) of the weak mixing angle, and the coefficients $C_G(f)$ are determined from the anomaly and are given in \eref{eq:CGcoeff} below; see~\cite{Low:2015qep, Berthier:2015vbb,  Bai:2016czm, Harigaya:2016eol, Draper:2016fsr, Howe:2016mfq} for examples that yield 750~GeV diphotons from these operators.   The coefficients are determined by both the superpotential
couplings, Eq.~(\ref{eq:Superpotential}), along with the axial transformation properties of the  
fermion fields $\Psi_f$ ($\psi_f$) in Dirac (Weyl) notation:
\begin{eqnarray} 
\partial_\mu J^{5 \mu} 
  \subset \sum_f 2\, i\, M_f\, \overline{\Psi}_f\, \gamma^5\, \Psi_f 
        = \sum_f 2\, i\, M_f\, \psi_f\, \psi_f + \text{h.c.} \, . 
\end{eqnarray}
Using the fact that the matrix element involving the gauge bosons $g_\mu$ and the axial current
$\big\langle g \,g| \partial_\mu J^{5 \mu} |0\big\rangle$
vanishes as the gauge boson energy goes to zero, the matrix element  
$\big\langle g\, g| \sum_f i\, M_f\, \psi_f\, \psi_f + \text{h.c.} |0\big\rangle$
can be related directly to the anomaly terms  in Eq.~(\ref{eq:anomalyterms}).  This in turn leads directly
to the two-body decay widths of $\Gamma(A_B \rightarrow g \,g)$  
that reproduce the full one-loop calculations
when the fermions are sufficiently heavy, see e.g. \cite{Howe:2016mfq} for explicit expressions.
Since these operators are generated by an axial anomaly, there are no contributions from the loops involving scalars.

\begin{figure}[t!]
\includegraphics[width = 0.8 \textwidth]{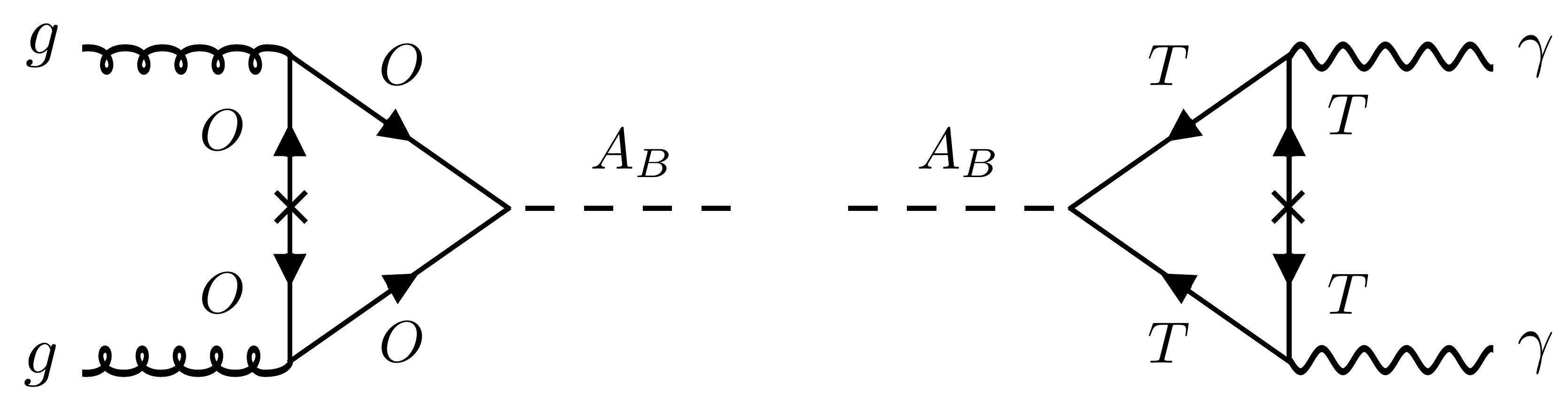}
\caption{The loop diagrams which yield the $A_B$ production via gluon fusion (left) and the subsequent decay to pairs of photons (right).}
\label{fig:LoopDiagrams}
\end{figure}

The superpotential coupling 
$S\, O \,O$ yields an effective interaction of $A_B$ with
gluons through loops of the octet fermions
as shown in Fig~\ref{fig:LoopDiagrams}, 
providing a production mechanism for an $s$-channel $A_B$.  
To understand the size
of the contribution, we will first consider a simplified model with 
a pure color octet fermion $\psi_O$.  
Both this state and the gluino must be
heavy to avoid direct search bounds.  Once $\psi_O$ is sufficiently heavier than
the pseudoscalar, $M_O \gg M_S$, it can be integrated out 
generating the gluon anomaly coefficient,  
\begin{eqnarray}
\frac{C_G(f)}{M_f} &=& \frac{N_c \,\lambda_{S O O}}{\sqrt{2}\, M_O} \, ,
\label{eq:CGcoeff}
\end{eqnarray}
where the number of colors $N_c$ arises from the Dynkin index of the octet. 
Using this coupling to calculate the width into gluons, 
we obtain the same result as the explicit loop calculation 
given in Eq.~(\ref{eq:gluonwidthexact}) from Appendix~\ref{app:Widths}.  
In fact, the effective
operator limit is reached relatively quickly, as the squared 
loop function given in Eq.~(\ref{eq:loopfunction}) that enters the width, 
$|\tau\, f(\tau)|^2 \simeq 1 + 2/(3\, \tau) + \mathcal{O}(1/\tau^2)$
is within $10\%$ of the exact result once $M_f \gtrsim 1.3 \, m_{A_B}$.

We have  both 
Dirac and Majorana masses producing  
gluino mass eigenstates $\widetilde{g}_{1,2}$ which are a mixture of   $\lambda_g$ and $\psi_O$ with mixing
angle $\theta_{\tilde{g}}$~\cite{Kribs:2013eua}:
\begin{eqnarray}
\left( \begin{array}{c} g_1 \\ g_2 \end{array} \right) &=&
\left( \begin{array}{cc} \cos\theta_{\tilde{g}} & -\sin\theta_{\tilde{g}} \\ 
                        \sin\theta_{\tilde{g}} &  \cos\theta_{\tilde{g}}
                        \end{array} \right) 
\left( \begin{array}{c} \psi_O \\ \lambda_g \end{array} \right) \ .
\end{eqnarray}   When both gluinos are at least
moderately heavy $m_{\tilde{g}_{1,2}} \gtrsim m_{A_B}$, 
the effective coupling of $A_B$ to gluons becomes
\begin{eqnarray}
\frac{C_G(O)}{M_f} &=& \frac{N_c\, \lambda_{S O O}}{\sqrt{2}}
\left( \frac{\cos^2\theta_{\tilde{g}}}{\big|M_{\tilde{g}_1}\big|} -
\frac{\sin^2\theta_{\tilde{g}}}{|M_{\tilde{g}_2}|} \right) \, .
\end{eqnarray}
Note that this coefficient vanishes in the pure Dirac
gluino case: $M_D \gg M_O, M_3$ implies 
$|M_{\tilde{g}_1}| \simeq |M_{\tilde{g}_2}|$ and
$\cos\theta_{\tilde{g}} \simeq \sin\theta_{\tilde{g}} \simeq 1/\sqrt{2}$. This is also clear in components since there is no coupling between $A_B$ and $\lambda_g$.
Using this expression for the couplings of the pseudoscalar
to the gluons, we obtain 
\begin{equation}
\Gamma(A_B\rightarrow g\,g) 
\simeq \frac{9\, \alpha_3^2\, \lambda_{SOO}^2 \,m_{A_B}^3}{16 \, \pi^3}
\left( \frac{\cos^2\theta_{\tilde{g}}}{|M_{\tilde{g}_1}|} -
\frac{\sin^2\theta_{\tilde{g}}}{|M_{\tilde{g}_2}|} \right)^2 \, .
\label{eq:GammaABgg}
\end{equation}
While this is a good approximation of the width into 
gluons, for completeness we use the full one-loop function 
from the Appendix in our numerical calculations.  

For much of the viable parameter space, the width to gluons will dominate the total width of $A_B$, implying that at least one 
gluino state cannot be arbitrarily decoupled, 
see Fig.~\ref{Fig:SignalCharginoGluino}.  As is clear both from the 
structure of the Feynman diagram and the
anomaly arguments, an odd number of Majorana mass insertions is required. 
A Dirac mass insertion connects the 
$O$ fermion with the usual MSSM gluino. However, only the 
$O$ component of the gluino couples to $A_B$, so an even number 
of Dirac mass insertions are needed  for a nonzero diagram ---  
either nonzero $M_3$ or $M_O$ are necessary to avoid 
the cancellation.  A large $M_3$  enhances the diagram, while 
large $M_O$ suppresses the coupling between $A_B$ and the gluons.  
We find that achieving large enough gluon coupling is feasible with 
a Dirac mass in the 1--4 TeV range and smaller Majorana masses. 
Because the large Dirac mass dominates the mass matrix, 
both gluinos can be heavy without unnaturally large renormalization 
of squark masses.

Similarly, in order to generate the decay to photons, there will 
be a contribution from both the electrically charged triplet fermions 
and the charged Higgsinos through the superpotential couplings
$S\, T\, T$ and $S\, H_u\, H_d$, respectively.  Here, the interactions of 
$A_B$ with electroweak gauge bosons are somewhat more 
complicated due to the mixing of the Higgsinos and 
triplet fermions into charginos and neutralinos. 
For heavy pure Higgsinos or triplet fermions, we verified that 
the decay widths obtained from the anomaly terms match the full-loop 
expressions given in the Appendix.  However, given that the
bounds on electroweakinos are considerably weaker than gluinos
we can take their masses to be much smaller, invalidating
the effective operator approach.  Moreover, the electroweakinos 
are generally far from pure states, due to electroweak 
symmetry breaking, Dirac masses, and Majorana masses.
The diagram showing $A_B \rightarrow \gamma\,\gamma$ 
involving the triplet fermions is given on the right of
Fig.~\ref{fig:LoopDiagrams}.  Loops of the charginos and neutralinos also lead
to significant contributions to the widths of $A_B$ into $\gamma\,Z$, 
$Z\,Z$, and $W^+\,W^-$, which are all included in the numerical results 
for the total width.  The full expressions are given in Appendix~\ref{app:Widths}.     

Unsurprisingly, we will show that at least one of these charginos must have a mass as light as possible without contributing to the on-shell decays of $A_B$, i.e., of $\mathcal{O}(400 \GeV)$, see Fig.~\ref{Fig:SignalCharginoGluino}. Direct searches for charginos and neutralinos do probe chargino masses of 400 GeV, even without slepton aided signatures. This places some constraint on the mass of the lightest neutralino; if it is heavier than $\sim180\gev$, there are no bounds on the chargino mass \cite{Aad:2014vma,Aad:2015jqa,Khachatryan:2014qwa,Khachatryan:2014mma}. With the need of light charginos for the diphoton width, they are likely to be observable via direct searches for these states in future runs of the LHC. For more on supersymmetric triplets producing large diphoton rates, see \cite{Arina:2014xya,Basak:2013eba,deBlas:2013epa,Delgado:2012sm,DiChiara:2008rg}.

We also consider potentially important tree-level contributions to the width of $A_B$.  The first two are the decays $A_B \rightarrow \widetilde{\chi}^0_i\,\widetilde{\chi}^0_j$ and $A_B \rightarrow \widetilde{\chi}^+_i\,\widetilde{\chi}^-_j$.  Whether or not these channels are open depends on the details of the spectrum.  Additionally, the rates vary depending on the admixtures of the neutralinos and charginos.  These effects are all included in the numerical results presented below; the explicit expressions used are given in Appendix~\ref{app:Widths}. We find the diphoton width can be large, even when decays to neutralinos are possible. Direct decays to charginos can also be accommodated, but this in general significantly reduces the diphoton rate. 

Tree-level decay to tops and bottoms are included and are possibly large, depending on the size of $\tan \beta$.  These decays constrain  the   mixing between the 750 GeV pseudoscalar and the standard $A^0$ pseudoscalar Higgs.  Specifically, there are three pseudoscalars in the model: $A_B$, $A_T$, and $A^0$.  The mixing between $A_B$ and $A^0$ could lead to the tree-level decay $A_B \rightarrow t\,\bar{t}$ and $A_B \rightarrow b\,\bar{b}$.  The width is given by
\begin{align}
\Gamma\big( A_B \rightarrow f\, \bar{f} \,\big) &= \sum_{t,b} N_c \frac{G_F\, m_f^2}{4\, \sqrt{2}\, \pi} \,g_{Aff}^2 \, \theta_{B0}^2\, m_{A_B} \left(1-4\frac{m^2_f}{m_{A_B}^2} \right)^{1/2} \notag\\
&\simeq \left[0.10 \text{ GeV}\,\left(\frac{20}{\tan\beta}\right)^2+9.4 \text{ GeV}\,\left(\frac{\tan\beta}{20}\right)^2 \right] \, \theta_{B0}^2,
\end{align}
where $g_{Aff} = \cot\beta$ for tops and $\tan \beta$ for bottoms, $G_F$ is the Fermi constant, $N_c = 3$ is the number of colors, $m_f$ is the fermion mass, and $\theta_{B0}$ is the mixing angle between $A_B$ and $A^0$.

For consistency with the $A_B \rightarrow \gamma \gamma$ rate, we require that the partial width to gluons dominates, i.e., $\Gamma(A_B\rightarrow t\,\bar{t}\,+b\,\bar{b})/\Gamma(A_B\rightarrow g\,g) \lesssim 0.1$. The typical size of the gluon width is $10^{-3} \gev$, which implies a limit on the mixing $\theta_{B0}^2 \lesssim 10^{-5}$.  In order to cast this in terms of a limit on the model parameters, we assume that the mixing is small such that the eaten Goldstone $G^0$ and the $A^0$ are mostly doublets.  Then in the mass eigenstate basis for $G^0$, $A^0$, and $A_B$, the mass mixing between $A^0$ and $A_B$ is
\begin{equation}
m_{A^0,A_B}^2 = \frac{v}{8}\, \text{Re}\left[ 4\,\sqrt{2}\,\lambda_{SHH} \,M_S-A_{\lambda_{SHH}} + 8 \,\lambda_{SHH} \,\kappa\, v_S - 4\,\sqrt{2}\,\lambda_{THH}\,\lambda_{STT}\, v_T \right],
\end{equation}
where $A_{\lambda_{SHH}}$ is the $A$-term for the $S\,H_u\,H_d$ scalars.  As discussed above, we are working in the region of parameter space where $v_S$ and $v_T$ can be neglected.  Given the expectation that $A$-terms are small in supersoft models, we can estimate the additional correction to this term expected from RGE running, where the $\beta$-functions were computed using the \texttt{SARAH} program~\cite{Staub:2010jh, Staub:2012pb, Staub:2013tta}.  Specifically, for reasonable benchmark values based on the results presented below ($\lambda_{SHH} = 0.3$, $M_2 \sim \text{few} \times \tev$, and running down from 100 TeV), we find that $A_{\lambda_{SHH}}\sim \text{few} \times 10 \gev$.  Therefore, the impact of these $A$-terms can be consistently neglected.  Finally, the constraint on the irreducible contribution is
\begin{equation}
\theta_{B0} \sim \frac{v}{\sqrt{2}} \frac{\lambda_{SHH}\, M_S}{m_{A^0}^2} \lesssim 3\times10^{-3} \quad\quad \Longrightarrow \quad \quad m_{A^0} \gtrsim 1800 \text{ GeV} \sqrt{\frac{\lambda_{SHH}}{0.3}} \,\sqrt{\frac{M_S}{200 \text{ GeV}}}.
\end{equation}
Given that naturalness considerations lead one to expect that $m_{A^0} \lesssim \text{few}\times \tev$; it is possible that $A_B$ decays to top or bottom quark pairs will be observable at the LHC.

These are all the effects that will be included in the numerics below in Sec.~\ref{sec:NumericalResults}. 
 Next we discuss the Standard Model--like Higgs boson mass, followed by the Landau poles due to the presence of large superpotential couplings, and finally the detailed results demonstrating the most viable regions of parameter space.

\section{Numerical Results}
\label{sec:NumericalResults}
As discussed above, the Dirac gaugino extension of the MSSM has all the ingredients to in principle produce the 750 GeV diphoton excess.  In this section, we will examine how well this model can do as a quantitative explanation.  Given that the best fit $\sigma \times \text{BR}_{\gamma\gamma} \simeq 3.8 \text{ fb}$ is somewhat larger than the typical expectation for a loop induced decay like the one we are relying on in this model (see Fig.~\ref{fig:LoopDiagrams}), it will be imperative to investigate what size  the couplings need to be.  As we will discuss in the following subsection, compatibility with the Higgs mass provides one set of constraints.  This will be followed by Sec.~\ref{sec:LandauPoles} which is devoted to understanding the correlation between large couplings and the presence of Landau poles.  We will end this section with a numerical exploration of the parameter space that is compatible with the diphoton excess.

\subsection{The Standard Model--like Higgs boson mass}
The Standard Model--like Higgs boson mass can be computed in this model:
\begin{align}
m_h^2 = m_Z^2\left[\cos^2 2\,\beta + \frac{2}{g_1^2+g_2^2}\left(\lambda_{SHH}^2+\frac{1}{2}\lambda_{THH}^2\right)\sin^2 2\,\beta\right] + \Delta_{m_h^2}^{(y_t)} + \Delta_{m_h^2}^{(\lambda)},
\label{eq:mhPhys}
\end{align}
where we have assumed the decoupling limit, $m_Z$ is the $Z$ boson mass, $\Delta_{m_h^2}^{(y_t)}$ is the correction from top/stops~\cite{Martin:1997ns}, and $\Delta_{m_h^2}^{(\lambda)}$ are the corrections proportional to $\lambda_{SHH}$ and $\lambda_{THH}$ from loops and higher dimension operators~\cite{Ellwanger:2005fh, Ender:2011qh, Benakli:2011kz, Benakli:2012cy}.  The $\Delta_{m_h^2}^{(\lambda)}$ contributions are computed about $v=0$, and include both the dominant one-loop effects along with the tree-level effects from integrating out all the heavy $S$ and $T$ states that couple to the Higgs~\cite{Benakli:2012cy}.  The explicit expressions used are given in Appendix~\ref{app:HiggsMassCorrections}, along with the detailed parameter choices.  The effects captured by $\Delta_{m_h^2}^{(\lambda)}$ are important away from small $\tan\beta$, where the $\sin 2\,\beta$ dependence of tree-level contribution implies a suppression.  

One property of Dirac gaugino models is that the $D$-term contribution to the Higgs mass is zero in the limit that the bino and wino masses are exactly Dirac, implying that the physical Higgs lies along the $D$-flat direction i.e., $\tan \beta =1$.  This is also the limit that enjoys the largest correction from the tree-level $\lambda$-dependent contribution to the physical Higgs mass as in \eref{eq:mhPhys}.  However, as we will argue below (and realize explicitly in the numerical explorations of the parameter space), nontrivial Majorana masses for the wino are required to explain the diphoton excess.  This implies that $\tan \beta \gtrsim 1$ and so we include the $SU(2)_L\times U(1)_Y$ $D$-term contribution, and will investigate a range of values for $\tan \beta$.

\begin{figure}[t]
\includegraphics[width=0.5\linewidth]{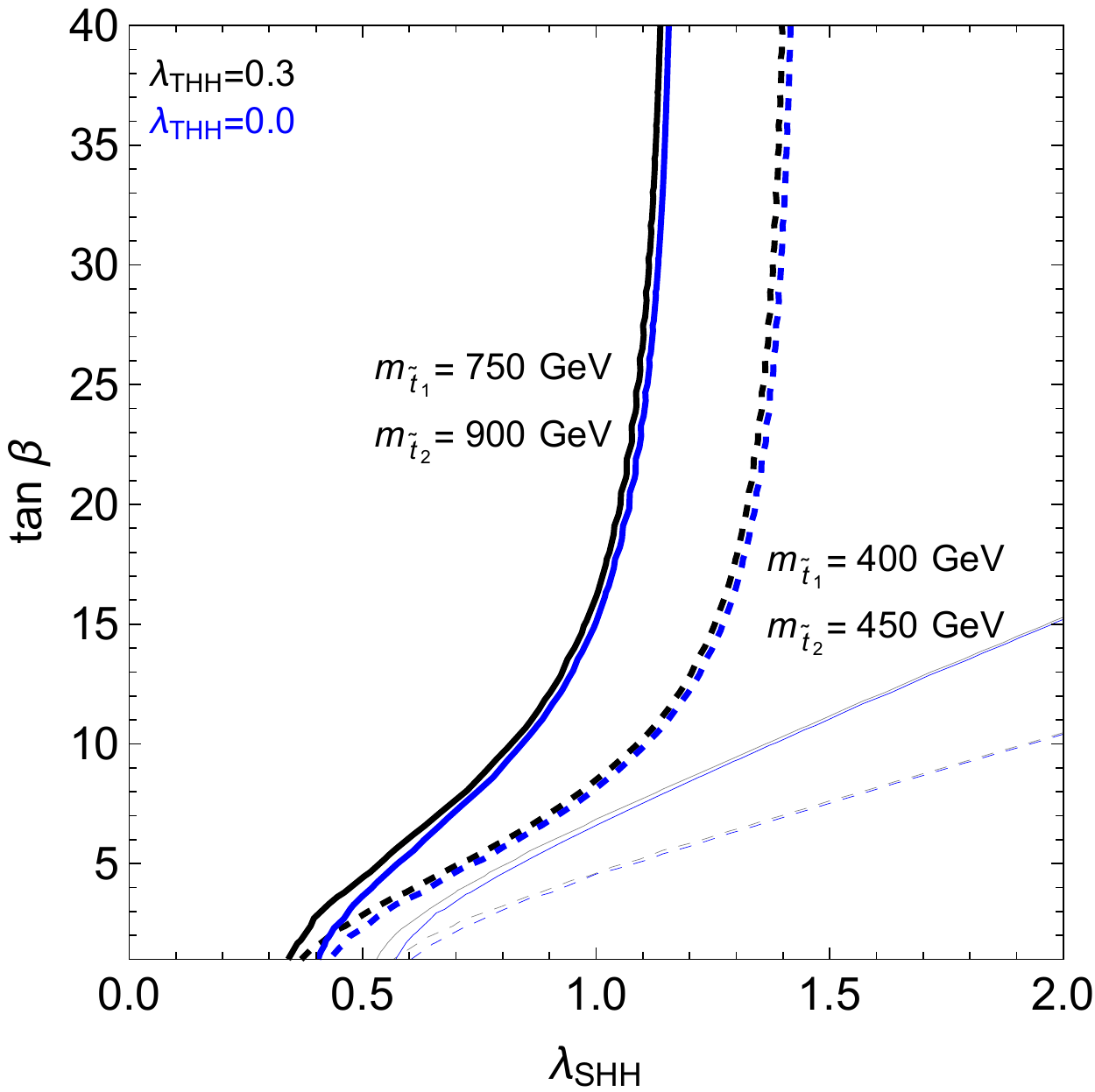}
\caption{Values of $\tan\beta$ and $\lambda_{SHH}$ which produce a 125 GeV Standard Model--like Higgs boson.  The decoupling limit has been assumed.  The black (blue) curves have $\lambda_{THH} = 0.3$ ($\lambda_{THH} = 0$).  Two benchmark choices for the stop masses have been chosen ``light stops'' $m_{\tilde{t}_1} = 750 \gev$ and  $m_{\tilde{t}_2} = 900 \gev$ for the solid lines, and ``very light stops'' $m_{\tilde{t}_1} = 400 \gev$ and  $m_{\tilde{t}_2} = 450 \gev$ for the dashed lines and the stop mixing $\theta_t=0$ for both.  Note that the second set of benchmark values implicitly assumes that the LHC limits can be avoided by e.g. having some compression between the stops and the lightest neutralino.  The thick curves are the full result for $m_h$.  The thin lines only include the tree-level contributions and the one-loop contributions from the stops, and neglect the loop contributions from the singlet and triplet couplings with the Higgs fields in order to demonstrate the impact of these higher order effects.}
\label{fig:HiggsMass}
\end{figure}

In Fig.~\ref{fig:HiggsMass}, we show the values of $\tan\beta$ and $\lambda_{SHH}$ which yield a Higgs mass of 125 GeV under a few different assumptions. The black  (blue) curves show the values for the triplet coupling $\lambda_{THH}=0.3$ ($\lambda_{THH}=0$). In our numerical investigations, we find that a large value of the Higgs-triplet coupling lowers the signal due to additional mixing between the charginos. For this reason, we take a relatively small value of $\lambda_{THH}=0.3$ for the numerical results presented below. 

For simplicity, we have taken the stop mixing angle $\theta_t = 0$.  We also provide two different benchmark choices for the stop masses, which in turn impact the value of $\Delta_{m_h^2}^{(y_t)}$. The dashed lines are relevant for ``very light stops'' ($m_{\tilde{t}_1} = 400 \gev$ and  $m_{\tilde{t}_2} = 450 \gev$), which can avoid detection if e.g. the spectrum is somewhat compressed or if $R$-parity violating decays obscure the signature sufficiently.  The solid lines are for more conventional values of the stop masses ($m_{\tilde{t}_1} = 750 \gev$ and  $m_{\tilde{t}_2} = 900 \gev$), which are just outside current limits~\cite{Aad:2015pfx, Aad:2014bva,Aad:2014kra,Aad:2014qaa,Aad:2014nra,Aad:2014mha,ATLAS:2014fka,Aad:2013gva, Khachatryan:2016xvy,Khachatryan:2016kdk,CMS:2016qtx,Khachatryan:2016zcu,CMS:2014dpa,Khachatryan:2014doa,Khachatryan:2015wza}.  

There can also be significant higher order corrections to the Higgs mass coming from the couplings with the triplet and the singlet, $\Delta_{m_h^2}^{(\lambda)}$ of Eq.~\eqref{eq:mhPhys}. These corrections depend on the detailed mass spectrum for all the components of the $S$ and $T$ superfields.  Note that custodial symmetry violations as measured by the $\rho$ parameter lead to a strong constraint on the vev of the triplet.  In the context of the full model, this can be interpreted as a lower bound on the physical mass of the triplet scalars $m_T \gtrsim 1.4\tev$~\cite{Baak:2012kk,Baak:2014ora,Beringer:1900zz, Davier:2010nc,ATLAS:2014wva,Aad:2014aba,CMS:2014ega,Benakli:2012cy,Alvarado:2015yna} (for the complete expression for the $\rho$ parameter in this model, see~\cite{Baak:2012kk}). For concreteness, we take the triplet scalar (pseudoscalar) mass to be 1.4 TeV (1.5 TeV). For compatibility with the diphoton signal, the bino-partner pseudoscalar is given a mass of 750 GeV, and for concreteness the bino-partner scalar mass is 900 GeV.  The curves in Fig.~\ref{fig:HiggsMass} shown with light, thin lines do not include these higher order corrections from the singlet and triplet. Since these corrections have a different dependence on $\tan\beta$ than the tree-level piece, we see that they have a dramatic impact on the parameter space.

Unsurprisingly, when the stops are near the uncompressed LHC limits, achieving a 125~GeV Higgs mass requires a smaller contribution from the supersymmetric couplings, and $\lambda_{SHH} \lesssim 1$. On the other hand, when the stops are very light, there is room for a large contribution from the superpotential, and  we obtain an approximate upper bound on the value for the singlet-Higgs coupling of $\lambda_{SHH} \lesssim 1.3$.  This impacts our interpretation of the allowed parameter space below.

\subsection{Landau poles}
\label{sec:LandauPoles}

In this section, we provide the correlation between the couplings evaluated at the TeV scale and the scale where Landau poles will appear using the low energy particle content of the model described in Sec.~\ref{sec:TheModel}.  Using the \texttt{SARAH} program~\cite{Staub:2010jh, Staub:2012pb, Staub:2013tta}, we have obtained the relevant $\beta$-functions for the couplings $\lambda_{SHH}$, $\lambda_{STT}$, $\lambda_{THH}$, $\lambda_{SOO}$, the gauge couplings, and the top Yukawa, see Appendix~\ref{app:RGEs} for the explicit expressions.  In order to explore the effects of the singlet-octet coupling, we choose three values for the $\lambda_{SOO}$ coupling, 0.3, 0.6, and 1.2.  As discussed above, we have fixed $\lambda_{THH}=0.3$. 

\begin{figure}[h!]
\includegraphics[width = \linewidth]{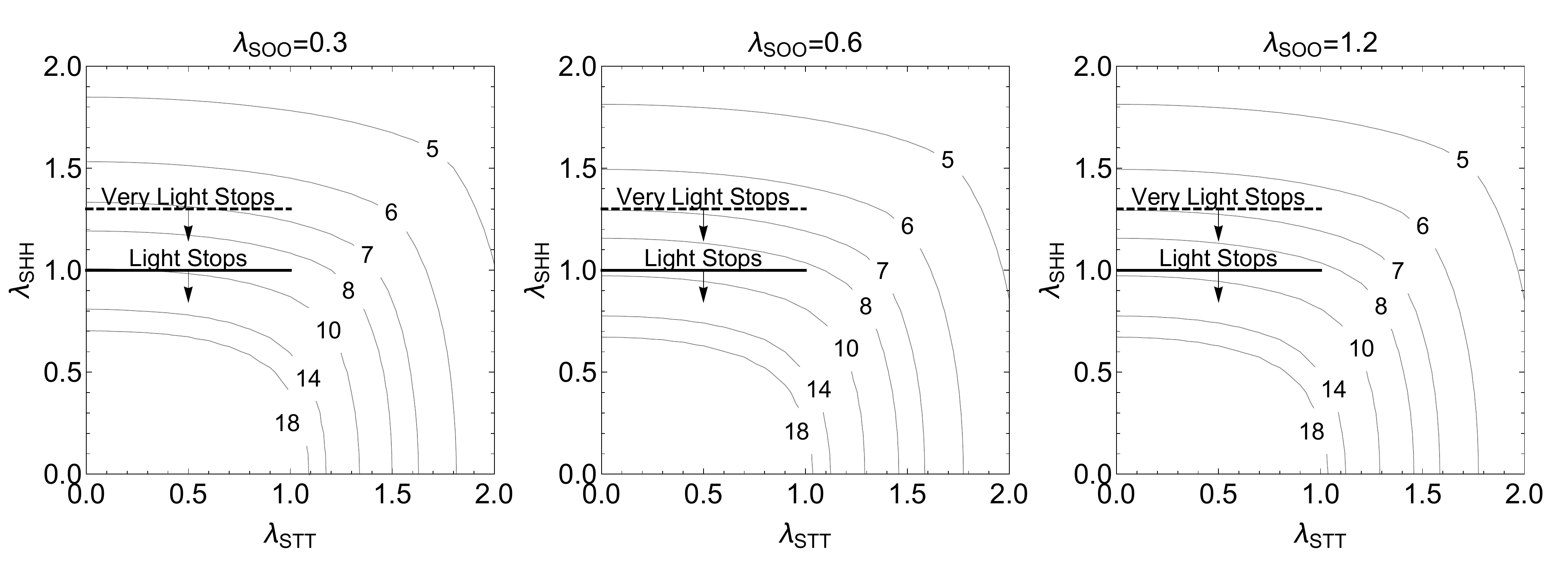}
\caption{Shown here are contours of $\log_{10}\Lambda_\text{Landau}/\text{GeV}$, the scale where at least one coupling becomes nonperturbative, in the $\lambda_{SHH}$ versus $\lambda_{STT}$ plane.  The (left, middle, right) panels are for $\lambda_{SOO} = (0.3, 0.6, 1.2)$.  The horizontal lines denote the upper bound on the coupling due to requiring compatibility with a 125 GeV Higgs mass:  $m_{\tilde{t}_1} = 750 \gev$ and  $m_{\tilde{t}_2} = 900 \gev$ yields the ``light stops'' solid line, and $m_{\tilde{t}_1} = 400 \gev$ and  $m_{\tilde{t}_2} = 450 \gev$ yields the  ``very light stops'' dashed line.  We have fixed $\lambda_{THH}=0.3$ in this figure.}
\label{fig:RGEs}
\end{figure}

The results are given in Fig.~\ref{fig:RGEs}, where we show contours of the approximate scale $\log_{10}\Lambda_\text{Landau}/\text{GeV}$, where a Landau pole would appear  in the $\lambda_{SHH}$ versus $\lambda_{STT}$ plane.  This assumes no new particles appear between the TeV scale and this new scale that would impact the running of these couplings. Note that in minimal Dirac gaugino models, gauge unification is absent.  However, additional particle content which would in principle have minimal impact on the RGE evolution of the superpotential couplings studied here can yield unification into e.g. $SU(3)^3$~\cite{Fox:2002bu}. Without the extra content, for a modest cutoff of $10^5 \text{ GeV}$, it is possible to realize $\lambda_{STT} \simeq 2$.  This is in part due to the slightly larger $SU(2)_L$ Casimir for the triplet superfield.  We also denote the region of the plot that is consistent with the Higgs mass, assuming the ``very light stop'' scenario with masses are around 400 GeV, or for the ``light stop'' scenario where their masses are around 750 GeV.  It is clear that a larger value of $\lambda_{SOO}$ implies a lower cutoff.   From the RGE point of view, it is also consistent to have a large value of $\lambda_{SHH}$, however this is disfavored by the requirement of a 125 GeV Standard Model--like Higgs mass unless the stops are light.  We will find below that it will be possible to generate the diphoton cross section for parameters in the ``light" stop range and with Landau poles that approach the Planck scale.

\subsection{Viable parameter space}
Finally, we  present the parameter space which is theoretically consistent and accommodates the 750 GeV diphoton excess. To see how the production cross section affects the signal, we performed detailed explorations of parameter space for a smaller (0.6) and larger (1.2) choice of the $\lambda_{SOO}$ coupling. For each of these, we then scan over values for the couplings responsible for the diphoton decay. We make a grid of $\lambda_{STT}$ and $\lambda_{SHH}$ in the range [0, 2] with a step size of 0.2. As discussed above, we keep $\lambda_{THH}$ fixed at 0.3.

For each grid point, models are generated using random values for the mass parameters (Dirac and Majorana) of the fermions of the singlet, triplet, octet, gauginos, and Higgsinos. The values of the mass parameters are chosen in the range  $[-3000 \text{ GeV}, 3000 \text{ GeV}]$. Motivated by Fig.~\ref{fig:HiggsMass}, we use $\tan\beta=20$, which allows us to reproduce the observed Higgs mass with larger values of the $\lambda_{SHH}$ coupling.  The gluino, chargino, and neutralino mass matrices are computed and diagonalized for each point in the scan. From these, the couplings between $A_{B}$ and the neutralinos, charginos, and gluinos are obtained. We then calculate the partial widths for $A_B\rightarrow \left\{ \gamma\, \gamma, \gamma\, Z, Z\,Z, W^+\, W^-, g\,g, \widetilde{\chi}^0_i \,\widetilde{\chi}^0_j, \widetilde{\chi}^{+}_i\, \widetilde{\chi}^-_j ,t\,\bar{t}, b\,\bar{b} \right\}$, see Appendix~\ref{app:Widths} for more details.\footnote{To simplify the computation of the widths for $\gamma\, Z, Z\,Z,$ and $ W^+ \,W^-$, we have taken the approximation that the chargino and neutralino couplings with the $W$ and $Z$ are vectorlike with a strength equal to the average of the actual left- and right-handed couplings. Over most of the relevant parameter space, the two couplings are within a few percent difference of each other.} The signal cross section is computed in the narrow width approximation as
\begin{eqnarray}
\sigma\big(p\,p\rightarrow A_B \rightarrow \gamma \,\gamma\big ) &=&\sigma\big(p\,p\rightarrow A_B\big) \,\text{BR}\big(A_B \rightarrow \gamma\,\gamma\big) \nonumber \\
&=& \frac{K\, c_{gg}}{m_{A_B}\, s}\Gamma\big(A_B \rightarrow g\,g\big) \frac{\Gamma\big(A_B \rightarrow \gamma\,\gamma\big)}{\Gamma_{\text{total}}}
\label{eqn:XS}
\end{eqnarray}
where $\sqrt{s}$ is the center of mass energy and $c_{gg}= 2137$ is the gluon partonic integral at $\sqrt{s}=13\tev$ \cite{Franceschini:2015kwy}. The next-to-next-to-leading-order $K$-factor for the production of a 750~GeV pseudoscalar is $K \simeq 3$~\cite{Harlander:2002vv, Howe:2016mfq}. The partial width to gluons dominates the total width for the most viable regions of parameter space.  

Taking naturalness as a guide, we are most interested in the region of parameter space where the supersoft and supersafe mechanisms are in effect. Specifically, naturalness favors a gluino mass which is  dominated by the Dirac term. This need not be the case for the electroweakinos since their impact on naturalness is mild.  Furthermore, the electroweak $D$-term contributions to the Higgs mass could be nontrivial if the bino and/or wino has a large Majorana mass.  Therefore, we only require the gluino to be Dirac-like in the figures presented below; this is imposed numerically by requiring that $M_3 < M_O^D$ and $M_O < M_O^D$.

Each of the two grids was sampled originally for around 1200 core hours. Every model which had Dirac-like gluinos and had the lightest neutralino lighter than the lightest chargino was saved. The goal was to get a clear picture of the maximum signal obtainable at each point of the couplings in the parameter space. We then took the seeds from the scan and found the model with the largest signal rate at each point in the grid. Another sampling for a total of around 500 core hours was preformed on both of the grids, only allowing the mass parameters to drift $\pm25\%$ of the values at the maximal point, while still maintaining the requirements of a Dirac gluino and neutralino lighter than the lightest chargino. When a signal rate larger than the previous maximum was found, that model became the new seed point. 

\begin{figure}[t!]
\includegraphics[width= \linewidth]{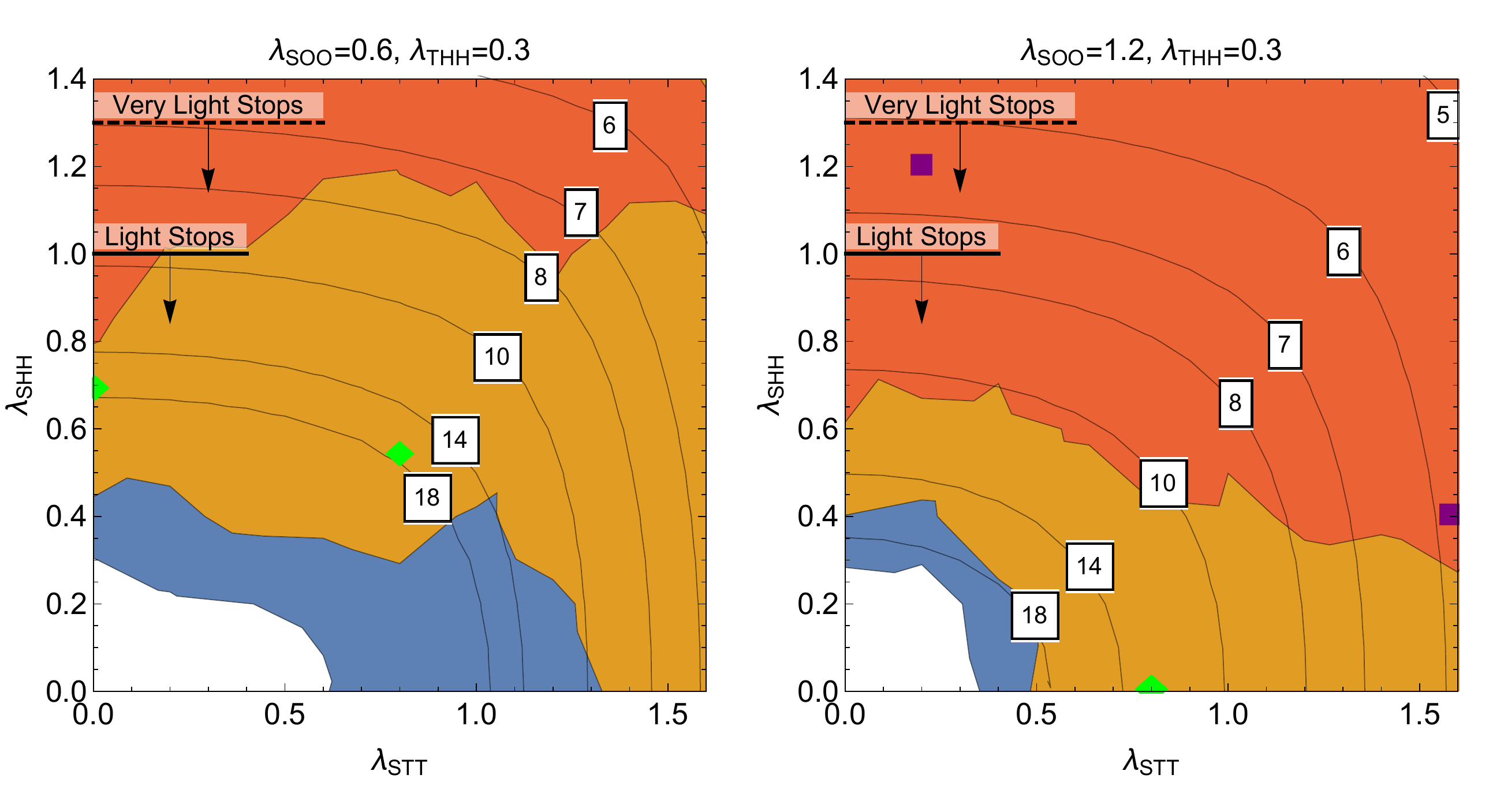}
\caption{The contours display $\log_{10}$ of the scale at which a supersymmetric coupling becomes nonperturbative. The left (right) panel has $\lambda_{SOO}=0.6$ ($\lambda_{SOO}=1.2$). The blue area is capable of producing a signal rate within the 2$\sigma$ best fit region defined by \cite{Buckley:2016mbr}. The orange region is capable of producing a signal within 1$\sigma$ of the best-fit value. The red region can produce a signal that is larger than $1\sigma$ \emph{above} the best fit signal, such that it is straightforward to find parameters which produce the observed rate. The purple boxes and green diamonds denote benchmarks that are studied in more detail in this section, the former are chosen to emphasize naturalness while the later have Landau poles which are postponed to near the GUT scale (and also have accompanying Fig.~\ref{Fig:SignalCharginoGluino}). The horizontal lines denote the upper bound on the coupling due to requiring compatibility with a 125 GeV Higgs mass:  $m_{\tilde{t}_1} = 750 \gev$ and  $m_{\tilde{t}_2} = 900 \gev$ yields the ``light stops'' solid line, and $m_{\tilde{t}_1} = 400 \gev$ and  $m_{\tilde{t}_2} = 450 \gev$ yields the  ``very light stops'' dashed line. We have fixed $\lambda_{THH}=0.3$ in this figure. The minimum gluino mass considered in these plots is 1.5 TeV. The shapes are not exact, with a grid step size of 0.2, and the value at each point coming from the maximization algorithm depending on random sampling.}
\label{Fig:ContoursGluino}
\end{figure}

Figure \ref{Fig:ContoursGluino} shows the contours of the signal rate obtained from this scanning method. In the white regions, the signal rate is always at least 2$\sigma$ below the best fit region defined by \cite{Buckley:2016mbr}. The blue regions have a maximal signal which lies between the 1 and 2$\sigma$ lower band of the best fit; the orange region's models have a maximum signal rate which is within 1$\sigma$ of the best fit; the red region has models which can produce a signal larger than 1$\sigma$ above the best fit --- in this region of parameter space, there is more freedom to choose the masses while still being consistent with the signal. Note that the jaggedness of these contours is not physical, both because the grid size is relatively large (0.2), and that the maximum signal rates are found from random scanning. The left and right panels show the small and large values for $\lambda_{SOO}$, respectively. As discussed earlier, this coupling affects the scale at which a Landau pole would appear. The contours (as before) show $\log_{10} \Lambda_{\text{Landau}}/\text{GeV}$. We have again marked the regions of the plot consistent with the Higgs mass given the stop mass assumptions as before.

In addition, we have imposed two phenomenological constraints on the grids. First, we demand that the lightest gluino be heavier than 1.5 TeV, which is roughly consistent with 13 TeV gluino searches with decays to first and second generation squarks \cite{ATLAS-CONF-2015-06, Aad:2016jxj}. It is worth acknowledging some tension with the latest results from the LHC, as the Higgs mass constrains us to have light stops. However, there is likely some level of compression between the stops and the neutralinos/charginos, along with some level of boost for the final states, complicating the searches. A detailed recasting of the limits for this model is beyond the scope of this paper and left for further analysis. As such, we use 1.5 TeV as the benchmark cutoff for the lightest gluino and leave a full exploration of the allowed parameter space to future work. Second, we only consider model points not explicitly excluded by 13 TeV $Z\,\gamma$ resonance searches \cite{ATLAS-CONF-2016-010, CMS-PAS-EXO-16-019}. The ATLAS analysis had a downward fluctuation which makes their exclusion more powerful than CMS, and so we use the reported ATLAS constraint at 750 GeV to place a cut of $\sigma\times \text{BR} \big( A_B \rightarrow Z\,\gamma \big) > 25$ fb. There is some tension in regions of parameter space where the triplet charginos are responsible for the diphoton signal~\cite{Howe:2016mfq}, see the ``Triplet" benchmark in Table~\ref{tab:BM}.

\begin{table}[t!]
{\footnotesize 
\renewcommand{\arraystretch}{1.13}
\setlength{\tabcolsep}{12pt}
\setlength{\arrayrulewidth}{.3mm}
\begin{center}
\begin{tabular}{c|c | ccc|cc}
\hline
& & Higgsino & Mixed & Triplet & Natural$_T$ & Natural$_H$ \\
\hline
 \parbox[t]{2mm}{\multirow{3}{*}{\rotatebox[origin=c]{90}{Couplings}}} & $\lambda_{SHH}$ & 0.7 & 0.55 & 0 & 0.4 & 1.2\\
&$\lambda_{STT}$ & 0 & 0.8 & 0.8 & 1.6 & 0.2\\
&$\lambda_{SOO}$ & 0.6 & 0.6 & 1.2  & 1.2 & 1.2\\
\hline
&$\Lambda_\text{Landau}$ [GeV]  & $10^{16}$ & $10^{17}$ &  $10^{13}$  & $10^{6}$ & $10^{6}$\\
 \hline
 \parbox[t]{2mm}{\multirow{10}{*}{\rotatebox[origin=c]{90}{Input Mass Parameters [GeV]}}} & $M_1$ & -3016 & -3016 & -459 & -2680 & -1020\\
& $M_2$ & 80 & -991 & -2281 & -2620 & -920\\
& $M_3$ &-1964 & -1964 & 2660 & -730 & 525\\
\cline{2-7}
& $M_{S}$ & 2263 &2260 & -654 & -1200 & 1330\\
& $M_{T}$ & -1526 & 1700 & -407 & -580 & -1370\\
& $M_{O}$ & 385 & 385 & 109 & 812 & -600\\
\cline{2-7}
& $M_{S}^D$ & 3110 & 3110 & -335 & -1520 & 1100\\
& $M_{T}^{D}$ & 4000 & 200 & 236 & 1700 & 3200\\
& $M_{O}^{D}$ & 1991 & 1991 & -2682 & 1450 & 1600\\
\cline{2-7}
& $\mu$ & -377 & 379 & -1100 & 800 & -390\\
\hline
 \parbox[t]{2mm}{\multirow{8}{*}{\rotatebox[origin=c]{90}{\quad Physical Masses [GeV]}}}  & ${\tilde{\chi}^{\pm}_1}$ & 378 & 376 & 376 & 375 &391 \\
& ${\tilde{\chi}^{\pm}_2}$ & 3360 & 1013 & 1100 & 805 & 2065\\
& ${\tilde{\chi}^0_1}$ & 375 & 375 & 207 & 241 & 380\\
& ${\tilde{\chi}^0_2}$ & 378 & 379 & 378 & 384 & 398\\
& ${\tilde{\chi}^0_3}$ & 3360 & 1012 & 903 & 802 & 1460\\
& ${\tilde{g}_1}$ & 1520 & 1520 & 1585 & 1600 & 1660\\
& ${\tilde{g}_2}$ & 3100 & 3100 & 4354 & 1680 & 1730\\
\hline
 \parbox[t]{2mm}{\multirow{12}{*}{\rotatebox[origin=c]{90}{Observables}}} & $\Gamma_{\text{tot}}$ [MeV] & 9.6 & 9.9 & 41.3 & 74.2 & 13.3\\
& $\Gamma\big(g\,g\big)$ [MeV] & 4.2 & 4.2 & 14.4 & 7.0 & 3.2\\
& $\Gamma\big(\gamma\,\gamma\big)$ [MeV]  & 0.40 & 0.37 & 0.57 & 1.7 & 0.78\\
& $\Gamma\big(\tilde{\chi}^0_1\,\tilde{\chi}^0_1\big)$ [MeV] & 0 & 0 & 1.9$\times10^{-6}$  &0.024 & 0\\
& $\Gamma\big(\tilde{\chi}^0_1\,\tilde{\chi}^0_2\big)$ [MeV] & 0 & 0 & 2.7$\times10^{-6}$ & 2.4$\times10^{-3}$ & 0\\
\cline{2-7}
& BR$\big(\gamma\,\gamma\big)$ & 0.041 & 0.037 &0.014 & 0.023 & 0.054\\
& BR$\big(g\,g\big)$ & 0.44 & 0.42 & 0.35 & 0.095 & 0.24\\
\cline{2-7}
& $\Gamma\big(W^+\,W^-\big)/\Gamma\big(\gamma\,\gamma\big)$ & 8.6 & 10.1 & 30.7 & 24.1 & 8.4 \\
& $\Gamma\big(Z\,Z\big)/\Gamma\big(\gamma\,\gamma\big)$ & 3.1 & 3.3 & 9.2 & 8.5 & 2.9\\
& $\Gamma\big(Z\,\gamma\big)/\Gamma\big(\gamma\,\gamma\big)$ & 0.71 & 1.1 & 6.1 & 5.5 & 0.69  \\
\cline{2-7}
& $\sigma\big(p\,p\rightarrow A_B\big)$ [fb] & 83 & 83 & 284 & 138 & 63\\
& $\sigma\times \text{BR}\big(\gamma\,\gamma\big)$  [fb] & 3.5 & 3.1 & 3.9 & 3.2 & 3.7\\
\hline
\end{tabular}
\end{center}
\caption{\setstretch{1} Benchmarks model points. The first three choices (corresponding the green diamonds in Fig.~\ref{Fig:ContoursGluino}) emphasize postponing the Landau pole to very high scales, and are further characterized by the identity of the lightest charginos: Higgsino-, Mixed-, and Triplet-like.  The two additional points (corresponding the purple squares in Fig.~\ref{Fig:ContoursGluino}) are chosen to emphasize  naturalness, and the subscript again refers to the identity of the lightest charginos.  The coupling $\lambda_{THH}=0.3$ for all model points. 
For reference, the best fit cross section is $\sigma_{\gamma\gamma} \simeq 3.8 \text{ fb}$~\cite{Buckley:2016mbr, BuckleyBlog}.}
\label{tab:BM}
}
\end{table}%
\afterpage{\FloatBarrier}

There is some difference in the shapes of the signal contours between the results in Fig.~\ref{Fig:ContoursGluino} for small (left) and large (right) $\lambda_{SOO}$ couplings. The main cause stems from the relative sizes of the electroweak decays.  For reference, we provide some explicit benchmark parameter points in Table~\ref{tab:BM}.  The first three, marked with green diamonds in Fig.~\ref{Fig:ContoursGluino}, postpone the Landau pole up to very high scales, while the last two, marked with purple squares in Fig.~\ref{Fig:ContoursGluino}, are chosen to emphasize naturalness considerations. First, we examine the case when the Higgsinos are responsible for the electroweak decays, which happens at low values of $\lambda_{STT}$. Since the Higgsinos are $SU(2)_L$ doublets, they have a smaller interaction with the $W$ and $Z$ bosons as compared to the triplets, while the couplings between the charginos and photons are only set by the electric charges.  Taken together, this yields some intuition for the trends seen in the Table.  The triplets have larger partial widths to electroweak final states while their width into diphotons is effectively the same size as for the Higgsinos.  This translates into the physics underlying the two panels of Fig.~\ref{Fig:ContoursGluino}.  When $\lambda_{SOO} = 0.6$, the electroweak final states have an easier time dominating over the width to gluons.  In order to achieve a large enough diphoton branching ratio requires a large coupling between $A_B$ and the charginos.  Since the width of the other electroweak decays is much larger for triplets than the Higgsinos, the triplets require a larger coupling.  For contrast, when the gluino coupling is large, the gluino contribution can easily dominate the total width. In this situation, the rate is set approximately only by the digluon and diphoton widths, causing the symmetric shape of the blue region in the right plot. As $\lambda_{STT}$ or $\lambda_{SHH}$ are further increased, the electroweak decays again become important in determining the signal rate.

The three benchmark points examined so far are perturbative up to very large energy scales, at least $10^{13}$ GeV and some as high as the GUT scale. However, a closer examination of the model parameters shows that there is some tension with naturalness considerations.  Despite having the Dirac mass for the gluinos larger than the other octet masses, large values of $M_3$ are still needed, which implies nontrivial corrections to the stop and Higgs soft mass parameters through the RGEs, and in addition the ``Triplet'' point has heavy Higgsinos. This tension can be decreased by relying on larger values of the couplings to generate the signal at expense of a lower cutoff.  For concreteness, we have also provided two additional benchmarks in Table~\ref{tab:BM}, where Natural$_T$ (Natural$_H$) are characterized by light tripletlike (Higgsino-like) charginos.  These models are marked with the purple squares in Fig.~\ref{Fig:ContoursGluino}, and lie in the red regions which can overproduce the diphoton signal.   Hence, we see that the gluinos can be made more Dirac-like, reducing the production cross section down to the observed value.  Furthermore, accommodating a 125 GeV Higgs is correlated with light stops which is in line with the expectation for a natural spectrum.  Note that Natural$_T$ has a relatively large $\mu$-term, and this would dominate the tuning unless the naive tree-level relation between the Higgs mass squared parameter and the Higgsino mass were modified~\cite{Cohen:2015ala, Nelson:2015cea}.  

\begin{figure}[t!]
\includegraphics[width=\linewidth]{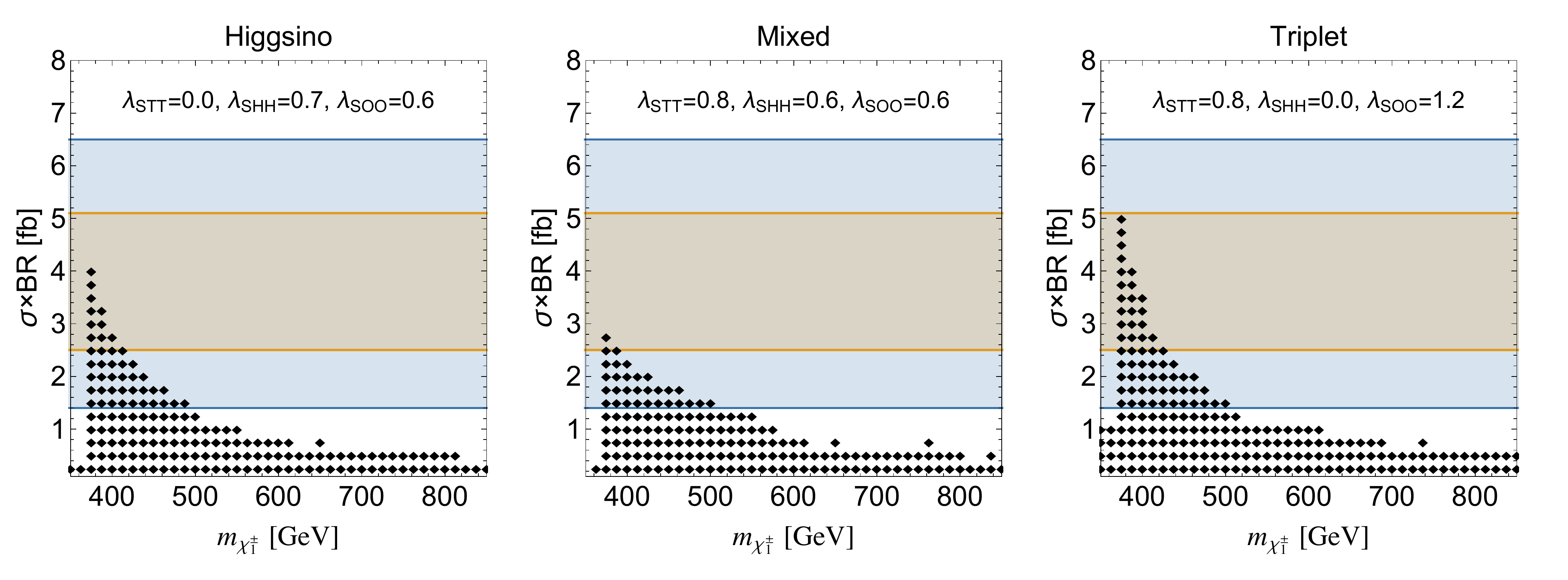}
\includegraphics[width= \linewidth]{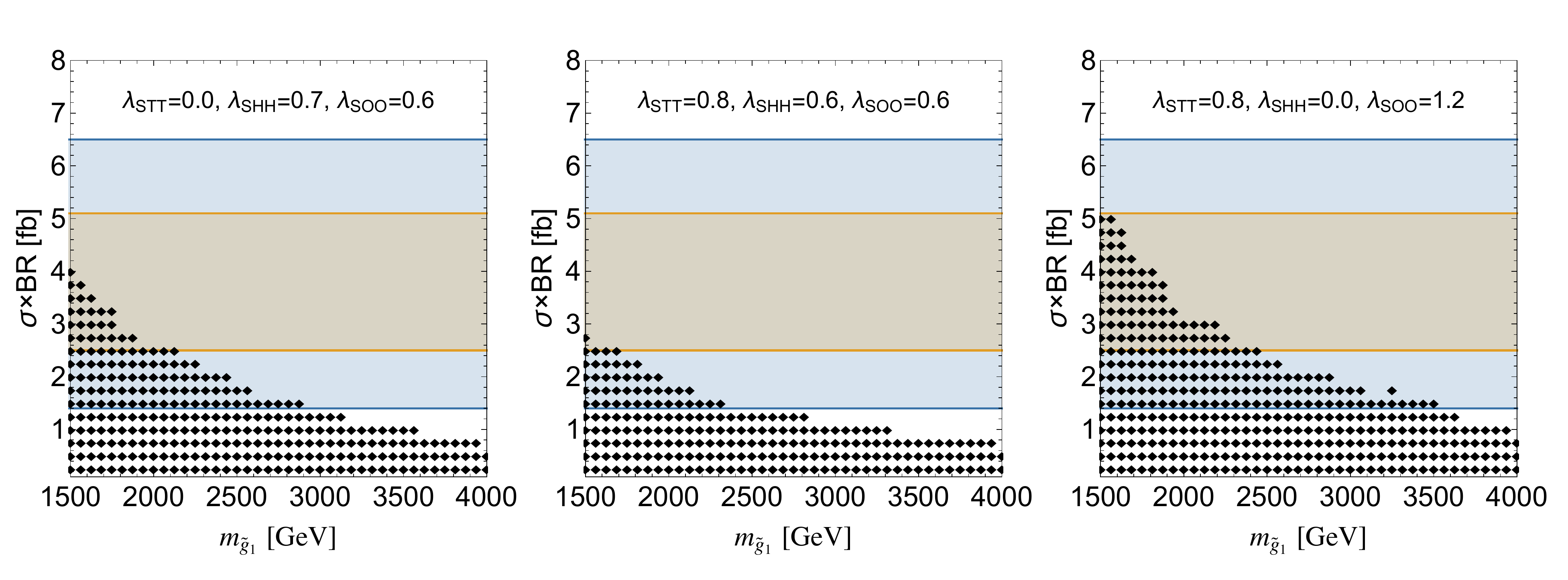}
\caption{Signal rate as a function of the lightest chargino (gluino) mass are given in the top (bottom) row. The left (right) panel has a small (large) value for the singlet-octet coupling.  The $1\sigma$ ($2\sigma$) region that is consistent with the diphoton excess~\cite{Buckley:2016mbr} is shaded in orange (blue).  We have fixed $\lambda_{THH}=0.3$ in this figure.  These points correspond to the green diamonds in Fig.~\ref{Fig:ContoursGluino}, and were chosen to realize models with a high scale Landau pole.  Heavier masses that are compatible with the signal can be realized at the expense of a lower cutoff.}
\label{Fig:SignalCharginoGluino}
\end{figure}

Finally, we show slices of the scan for fixed values of the couplings as a function of the mass of the lightest chargino or gluino in Fig.~\ref{Fig:SignalCharginoGluino}. The couplings chosen again correspond to the green diamonds of Fig.~\ref{Fig:ContoursGluino}. Rather than do a scatter plot of the many models examined, we instead make a grid over the signal rate and the mass eigenstate, if a model falls with the grid point, it is filled in. Figure \ref{Fig:SignalCharginoGluino} shows the signal rate plotted against the lightest chargino mass in the top row, where the orange (blue) areas show the 1$\sigma$ (2$\sigma$) bands.  Note that these couplings emphasize a high scale Landau pole.  All of the coupling choices are able to generate a signal within $1\sigma$ of the best fit value, but they do need the charginos to be near threshold. It is therefore reasonable to expect that the lightest chargino in this model should be $\lesssim 500 \gev$. While this does provide hope that the next run of the LHC could find the chargino, referring back to Table \ref{tab:BM}, the lightest neutralino (if it is not singletlike) tends to be nearly degenerate with the chargino. This compressed spectrum could make detection of the charginos very challenging.  The bottom row of Fig.~\ref{Fig:SignalCharginoGluino} gives the diphoton cross section as a function of the lightest gluino mass.  Here we see that the lightest gluino should be within reach of the LHC, for the majority of the parameter space.  However, this is strongly tied to the desire for perturbative couplings, and if one is willing to lower the scale of the cutoff, then the lightest gluino mass can reach the many TeV range while still being compatible with the diphoton signal.


\section{Conclusions}
\label{sec:Conclusions}
Motivated by the recent announcement of an excess in the diphoton invariant mass spectrum near 750 GeV, we have investigated the possibility that this signal is the first sign of supersymmetry.  The naturalness tensions of the MSSM from the Higgs mass and  direct searches for superpartners drive interest in  extensions of the MSSM.  Models where the gauginos have adjoint partners with Dirac and Majorana masses have all the ingredients to produce 750 GeV diphotons.  The   resonance is the pseudoscalar contained within the chiral superfield partner of the bino.  Its couplings to the gluino partner provides a production mechanism via gluon fusion while its interactions with the wino partners leads to a diphoton decay.  Furthermore, the supersoftness and supersafeness of the  model  allows it to be minimally tuned even in the presence of modern LHC limits.

We explored a number of interesting constraints on the parameter space.  Compatibility with a Standard Model--like Higgs boson at 125 GeV imposes an upper bound on the coupling between the bino partner and the Higgs superfields.  Furthermore, imposing perturbativity on the couplings  means that they cannot be arbitrarily large.  It is possible to reconcile both of these requirements while also accommodating the excess.  While this does not necessarily imply that gluinos will be visible at the LHC (although they would be accessible by a future proton collider), some of the charginos and neutralinos must be below about 500 GeV.  Furthermore, many other decay modes of the 750 GeV state should be observable, including other electroweak channels such as $\gamma\,Z^0$, $Z\,Z$, $W^+\,W^-$, along with decays to  $t\,\bar{t}$ and/or $b\,\bar{b}$.  

The possibility of an early discovery in diphotons at LHC13 is extremely exciting.  This work places it in the context of a complete TeV scale framework, which leads to many additional predictions for the LHC and future colliders.  If this signal persists, then we can expect a very rich program of new discoveries as more data is collected, leading to a revolution in the way we view the laws of nature.

\section*{Acknowledgements}
We thank Spencer Chang, Nathaniel Craig, Howard Haber, and Simon Knapen for useful conversations.  
T.C. is supported by an LHC Theory Initiative Postdoctoral Fellowship, under the National Science Foundation Grant No. PHY-0969510.  T.C. also thanks the KITP at UCSB, where this work was competed and the related support from the National Science Foundation under Grant No. NSF PHY11-25915.  This work was supported in part by the U.S. Department of Energy under Grants No. DE-SC0011640 (G.D.K. and B.O.) and No. DE-SC0011637 (A.E.N.). We also acknowledge the use of computational resources provided by the ACISS supercomputer at the University of Oregon, Grant No. OCI-0960354. 

\appendix
\section*{Appendix}
\section{Partial widths}
\label{app:Widths}

In this Appendix, we provide the relevant partial widths of $A_B$ into neutralinos, charginos, photons, and gluons. Starting with the neutralino mass matrix $\mathbf{M}_{\tilde{\psi}^0}$ in the gauge eigenstate basis $\psi^0 = \begin{pmatrix}  \widetilde{B}, \widetilde{W}^0, \widetilde{H}_d^0, \widetilde{H}_u^0, \widetilde{S}, \widetilde{T}^0 \end{pmatrix}$, we then diagonalize this matrix to derive the mixing matrix $Z_{ij}$:
\begin{equation}
\mathbf{M}_{0}^{\text{diagonal}} = Z^*\, \mathbf{M}_{\tilde{\psi}^0} \,Z^{-1}.
\end{equation}
The interaction eigenstates can then be written in terms of the mass eigenstates using $\psi^0_i = Z^{*}_{ji}\,\widetilde{\chi}^0_j$. The couplings of $A_B$ to the neutalinos is then given by
\begin{equation}
\Lc_{A_B\tilde{\chi}^0\tilde{\chi}^0} = A_B \,\overline{\widetilde{\chi}}^0_i \left( P_L\, g_L + P_R \,g_R \right) \widetilde{\chi}^0_j\,,
\end{equation}
where
\begin{eqnarray}
g_L &= &\frac{-i\, \lambda_{SHH}}{\sqrt{2}}Z^*_{i3}\,Z^*_{j4} + \frac{-i\, \lambda_{SHH}}{\sqrt{2}}Z^*_{i4}\,Z^*_{j3}+i\,\sqrt{2} \kappa Z^*_{i5}\, Z^*_{j5} +\frac{i\, \lambda_{STT}}{\sqrt{2}} Z^*_{i6}\,Z^*_{j6} \nonumber \\
g_R &=& \frac{i\, \lambda_{SHH}}{\sqrt{2}}Z_{i3}\,Z_{j4} + \frac{i\, \lambda_{SHH}}{\sqrt{2}}Z_{i4}\,Z_{j3} -i\,\sqrt{2} \kappa Z_{i5} \,Z_{j5}+\frac{-i\, \lambda_{STT}}{\sqrt{2}} Z_{i6}\,Z_{j6} = g_L^*\,.
\label{eqn:neutcoups}
\end{eqnarray}
Next we do the same procedure for the charginos with the mass matrix in the $\begin{pmatrix} \widetilde{W}^+ & \widetilde{H}_u^+ & \widetilde{T}^+ \end{pmatrix}\Big/\begin{pmatrix} \widetilde{W}^- & \widetilde{H}_d^- & \widetilde{T}^- \end{pmatrix}$ basis. This mass eigenstates are found in the usual way
\begin{equation}
\begin{pmatrix} \widetilde{C}_1^+ \\ \widetilde{C}_2^+ \\ \widetilde{C}_3^+ \end{pmatrix} = V \begin{pmatrix} \widetilde{W}^+ \\ \widetilde{H}_u^+ \\ \widetilde{T}^+ \end{pmatrix}\quad\quad \text{and} \quad \quad 
\begin{pmatrix} \widetilde{C}_1^- \\ \widetilde{C}_2^- \\ \widetilde{C}_3^- \end{pmatrix} = U \begin{pmatrix} \widetilde{W}^- \\ \widetilde{H}_d^- \\ \widetilde{T}^- \end{pmatrix}\,,
\end{equation}
where $U$ and $V$ are defined to diagonalize the chargino mass matrix, yielding real, positive masses with
\begin{equation}
\mathbf{M}_+^{\text{diagonal}} = U^*\, \mathbf{X} \,V^{-1} .
\end{equation}
The couplings of the charginos to $A_B$ can then be written as
\begin{equation}
\Lc_{A_B\tilde{\chi}^+\tilde{\chi}^-} = A_B\, \overline{\widetilde{\chi}}^-_i \left(P_L\, g_L + P_R \,g_R \right) \widetilde{\chi}^+_j
\end{equation}
with
\begin{eqnarray}
g_L &=& \frac{i\,\lambda_{SHH}}{\sqrt{2}} U^*_{i2}\, V^*_{j2} + \frac{i\, \lambda_{STT}}{\sqrt{2}} U^*_{i3}\,V^*_{j3} \nonumber \\
g_R &=& -\frac{i\, \lambda_{SHH}}{\sqrt{2}} V_{i2}\,U_{j2} - \frac{i\, \lambda_{STT}}{\sqrt{2}} V_{i3}\,U_{j3}.
\label{eqn:CharginoCouplings}
\end{eqnarray}
Having obtained the couplings, the decays to charginos or neutralinos are given by
\begin{align}
\Gamma(A_B \rightarrow \widetilde{\chi}_i \overline{\widetilde{\chi}}_j) &= \frac{m_{A_B}}{16\,\pi} \frac{1}{1+\delta_{ij}} \left[ \left(1-\frac{m_i^2}{m_{A_B}^2}-\frac{m_j^2}{m_{A_B}^2}\right)^2-\frac{4\,m_i^2\,m_j^2}{m_{A_B}^2}\right]^{1/2} \nonumber \\
&\times \left\{  \left(g_L \,g^*_L + g_R\, g^*_R\right) \left(1-\frac{m_i^2}{m_{A_B}^2}-\frac{m_j^2}{m_{A_B}^2}\right)-2\big( g_L\, g_R^*+g_L^*\, g_R\big) \frac{m_i \,m_j}{m_{A_B}^2 }\right\}.
\end{align}
where $\delta_{ij}=0$ except for the case of identical Majorana neutralinos.
The diphoton decays are given by
\begin{equation}
\Gamma(A_B \rightarrow \gamma\, \gamma) = \frac{\alpha^2\, m_{A_B}}{16\, \pi^3} \left| \sum_f N_c\, Q_f^2\, \lambda_{A_B f f} \,\sqrt{\tau}\, f(\tau) \right|^2 .
\label{eq:photonwidthexact}
\end{equation}
Only charginos run in the loop, so the number of colors $N_C$ and the charge of the fermion $Q_f$ are both equal to one. The coupling $\lambda_{A_B ff}$ can be inferred from Eq.~\eqref{eqn:CharginoCouplings} using $i=j$. We define $\tau=4\,m_f^2/m^2_{A_B}$. The loop function $f(\tau)$ is given by
\begin{equation}
f(\tau) = \left\{ \begin{matrix}\arcsin^2 \frac{1}{\sqrt{\tau}} &\tau \ge 1\\ -\frac{1}{4} \left[\log \frac{1+\sqrt{1-\tau}}{1-\sqrt{1-\tau}}-i\, \pi \right]^2 & \tau< 1.\end{matrix} \right.
\label{eq:loopfunction}
\end{equation}
For the decays to gluons, the partial width is given by
\begin{equation}
\Gamma(A_B \rightarrow g\,g) = \frac{\alpha_s^2\, m_{A_B}}{2 \,\pi^3} \left| \sum_f C(r)\, \lambda_{A_B ff}\, \sqrt{\tau}\, f(\tau) \right|^2~,
\label{eq:gluonwidthexact}
\end{equation}
where $\tau$ and $f(\tau)$ have the same meaning as before. The only particle running in the loop is the octet, although there are two physical gluino states. $C(r)$ is the Dyknin index of the representation running in the loop [defined as $\text{tr}[t^a_r t^b_r] = C(r)\delta^{ab}$], which is 3 for a color octet. The coupling is $\lambda_{SOO}/\sqrt{2}$ times the rotation angle derived from the gluino mass matrix.

We also provide approximate formulas for loop induced decays into the other electroweak vector bosons. Beyond the coupling of $A_B$ to the charginos and neutralinos shown in Eqs.~\eqref{eqn:neutcoups} and \eqref{eqn:CharginoCouplings} the couplings of the electroweakinos with the $W$ and $Z$ bosons are necessary as well. The charged current interactions are given by
\begin{equation}
\Lc \supset \widetilde{\chi}^0_j \,i\, \gamma^{\mu}\, W^-_{\mu} \left(g^L_{W^-\chi^0_j \chi^+_k} P_L + g^R_{W^-\chi^0_j \chi^+_k} P_R\right) \bar{\widetilde{\chi}}^+_k
\end{equation}
with the left- and right-handed couplings
\begin{eqnarray}
g^L_{W^-\chi^0_j \chi^+_k} &=& g \left( N_{j 2}\,V^*_{k1} +N_{j 6}\,V^*_{k3} -\frac{1}{\sqrt{2}} N_{j 4}\,V^*_{k2} \right) \nonumber \\
g^R_{W^-\chi^0_j \chi^+_k} &=& g \left( N^*_{j 2}\,U_{k1} +N^*_{j 6}\,U_{k3} +\frac{1}{\sqrt{2}} N^*_{j 3}\,U_{k2} \right).
\end{eqnarray}
The $Z$ couples to both charginos and the neutralinos separately:
\begin{align}
\Lc \supset &\, \widetilde{\chi}^+_m\, i\, \gamma^{\mu}\, Z_{\mu} \left(O^L_{Z \chi^+_m\chi^-_k}\, P_L + O^R_{Z \chi^+_m\chi^-_k} \,P_R\right) \bar{\widetilde{\chi}}^-_k \notag\\
& + \widetilde{\chi}^0_{\ell}\, i\, \gamma^{\mu}\, Z_{\mu} \left(g^L_{Z \chi^0_{\ell} \chi^0_n}\, P_L +g^R_{Z \chi^0_{\ell} \chi^0_n} \,P_R\right) \bar{\widetilde{\chi}}^0_n
\end{align}
with the couplings given by
\begin{eqnarray}
O^L_{Z \chi^+_m\chi^-_k} &=& \frac{g}{c_w} \left( - V_{m1}\, V^*_{k1} - V_{m3}\,V^*_{k3} - \frac{1}{2} V_{m2} \,V^*_{k2} + \delta_{mk} \,s_w^2 \right) \nonumber \\
O^R_{Z \chi^+_m\chi^-_k} &=& \frac{g}{c_w} \left( - U_{k1}\, U^*_{m1} - U_{k3}\,U^*_{m3} - \frac{1}{2} U_{k2}\, U^*_{m2} + \delta_{mk}\, s_w^2\right) 
\end{eqnarray}
and
\begin{eqnarray}
g^L_{Z \chi^0_{\ell} \chi^0_n} &=& \frac{g}{2 c_w} \Big(-N_{\ell3} \,N^*_{n3} + N_{\ell4} \,N_{n4}^* \Big) \nonumber \\
g^R_{Z \chi^0_{\ell} \chi^0_n} &=& -{g^{L}}^*_{Z \chi^0_{n} \chi^0_\ell}.
\end{eqnarray}

Unique left- and right-handed couplings greatly complicate the computation of the matrix element for the decay of $A_B \rightarrow V\,V$, including the presence of intermediate divergences that cancel when all diagrams have been summed over.  However, if the couplings are vectorlike, the expressions are finite at each step leading to significant simplifications. In the regions of parameter space which are interesting for generating the observed excess, the difference of the left- and right-handed couplings is small. In our calculations, we assume vectorlike couplings chosen to be the average strength of the left- and right-handed pieces.

The matrix element for a given diagram in this approximation is then given by
\begin{align}
\mathcal{M} = &\, g_{A_B f_1 f_3}\,\times\, g_{V_2 f_1 f_2}\,\times \, g_{V_1 f_2 f_3}\,\frac{1}{4 \,\pi^2}\, \varepsilon^{\mu\nu \alpha \beta}\, p_1^{\alpha} \, p_2^{\beta}\, \epsilon^{*}_{\mu}(p_1)\,\epsilon^{*}_{\nu}(p_2) \notag\\
& \times \int_0^1 \text{d}y \int_0^{1-y} \text{d}z \,\frac{m_{f_1}(y+z-1)-m_{f_2}\,y-m_{f_3}\,z}{\Delta}\,,
\end{align}
with $\Delta$ defined as
\begin{equation}
\Delta = m_{f_1}^2 (1-z-y)+m_{f_2}^2\, y + m_{f_3}^2\, z - y\,z \,m_{V_1}^2-y\,(1-y-z)\, m_{V_2}^2 -z\,(1-y-z)\,m_{A_B}^2.
\end{equation}
Define the generic integral
\begin{equation}
\mathcal{I}\big(m_{f_1},m_{f_2},m_{f_3},m_{V_1},m_{V_2}\big) = \int_0^1 \text{d}y \int_0^{1-y} dz \frac{m_{f_1}(y+z-1)-m_{f_2}y-m_{f_3}z}{\Delta}
\end{equation}
Squaring the matrix element and summing over the polarizations simplifies to
\begin{align}
\sum_{\text{polarizations}} \left| \mathcal{M} \right|^2 = &\,\frac{1}{16\, \pi^4} \left(2\,(p_1 \cdot p_2)^2 -p_1^2 \,p_2^2 \right)\notag\\ 
&\times \Big| g_{A_B f_1 f_3}\,\times\, g_{V_2 f_1 f_2}\,\times\, g_{V_1 f_2 f_3}~\mathcal{I}(m_{f_1},m_{f_2},m_{f_3},m_{V_1},m_{V_2}) \Big|^2.
\end{align}
The final momentums $p_1$ and $p_2$ depend on the masses of the final state vectors. Finally, this leads to the partial widths for the vector bosons. For $\gamma \,Z$,
\begin{align}
\Gamma\big(A_B \rightarrow \gamma\, Z\big) =&\, \frac{\alpha}{8 \,\pi^4} \frac{(m_{A_B}^2-m_Z^2)^3}{m_{A_B}^3} \notag\\
&\times \left| \sum_{i=1}^3\sum_{k=1}^3 g_{A_B \chi^+_i \chi^-_k} \,\times\, g_{Z \chi^+_k \chi^-_i}  ~\mathcal{I}(m_{\chi^{\pm}_i},m_{\chi^{\pm}_k},m_{\chi^{\pm}_k},0,m_Z) \right|^2.
\end{align}
The $Z\,Z$ decays have contributions from both the charginos and the neutralinos:
\begin{equation}
\Gamma\big(A_B \rightarrow Z\, Z\big) = \frac{(m_{A_B}^2-4\,m_Z^2)\sqrt{m_{A_B}^2-4\,m_{A_B}^2 \,m_Z^2}}{256 \,\pi^5\, m_{A_B}}\, \Big|\mathcal{M}_{\text{neutralinos}} + \mathcal{M}_{\text{charginos}} \Big|^2~,
\end{equation}
where
\begin{equation}
\begin{aligned}
\mathcal{M}_{\text{neutralinos}} =& \frac{1}{2} \sum_{i=1}^6 \sum_{j=1}^6 \sum_{k=1}^6 g_{A_B \chi^0_i \chi^0_k} \,\times\,g_{Z \chi^0_i \chi^0_j} \,\times\, g_{Z \chi^0_j \chi^0_k} ~\mathcal{I}\big(m_{\chi^{0}_i},m_{\chi^{0}_j},m_{\chi^{0}_k},m_Z,m_Z\big)\,,
\end{aligned}
\end{equation}
and
\begin{equation}
\mathcal{M}_{\text{charginos}} = \sum_{i=1}^3\sum_{j=1}^3 \sum_{k=1}^3 g_{A_B \chi^{+}_i \chi^{-}_k}\,\times\, g_{Z \chi^+_j \chi^-_i} \,\times\,g_{Z \chi^+_k \chi^-_j} ~ \mathcal{I}(m_{\chi^{\pm}_i},m_{\chi^{\pm}_j},m_{\chi^{\pm}_k},m_Z,m_Z)\,.
\end{equation}
The partial width for $W^+W^-$ is
\begin{equation}
\Gamma\big(A_B \rightarrow W^+\, W^-\big) = \frac{(m_{A_B}^2-4 \,m_W^2)\sqrt{m_{A_B}^2-4\,m_{A_B}^2\, m_W^2}}{128\, \pi^5\, m_{A_B}} \,\Big|\mathcal{M}_{\text{neutralino}}^{WW} +\mathcal{M}_{\text{chargino}}^{WW} \Big|^2\,,
\end{equation}
where
\begin{equation}
\mathcal{M}_{\text{chargino}}^{WW} = \sum_{i=1}^3 \sum_{k=1}^3 \sum_{j=1}^6   g_{A_B \chi^+_i \chi^-_k}\,\times\, g_{W^+ \chi^-_i \chi^0_j} \,\times\,g_{W^- \chi^+_k \chi^0_j} ~ \mathcal{I}\big(m_{\chi^{\pm}_i},m_{\chi^{0}_j},m_{\chi^{\pm}_k},m_W,m_W\big)\,,
\end{equation}
and
\begin{equation}
\!\!\mathcal{M}_{\text{neutralino}}^{WW} = \sum_{i=1}^6 \sum_{k=1}^6 \sum_{j=1}^3 ~ g_{A_B \chi^0_i \chi^0_k}\,\times\, g_{W^+ \chi^-_j\chi^0_i}\,\times\, g_{W^- \chi^+_j \chi^0_k} ~ \mathcal{I}\big(m_{\chi^{0}_i},m_{\chi^{\pm}_j},m_{\chi^{0}_k},m_W,m_W\big).
\end{equation}

\section{Renormalization group equations}
\label{app:RGEs}
The one-loop RGEs for superpotential couplings are given by 
\begin{align}
16\, \pi^2\, \frac{\text{d} g_1(t)}{\text{d}t} &= \frac{33}{5}\, g_1^3  \\
16 \,\pi^2\, \frac{\text{d} g_2(t)}{\text{d}t} &= 3 \,g_2^3  \\
16\, \pi^2\, \frac{\text{d} g_3(t)}{\text{d}t} &= 0  \\
%
16\, \pi^2\, \frac{\text{d} Y_t(t)}{\text{d}t} &= Y_t\left(6 \,Y_t^2 -\frac{13}{15}\, g_1^2- 3\, g_2^2 - \frac{16}{3}\, g_3^2 + \lambda_{SHH}^2 + \frac{3}{2}\, \lambda_{THH}^2\right)   \\
%
%
16\, \pi^2\, \frac{\text{d} \kappa(t)}{\text{d}t} &= \frac{3}{2}\,\kappa \left(3 \,\lambda_{STT}^2 + 4\, \kappa^2 + 4\, \lambda_{SHH}^2 + 8\, \lambda_{SOO}^2 \right) \\
16\, \pi^2\, \frac{\text{d} \lambda_{SHH}(t)}{\text{d}t} &= \lambda_{SHH} \left( -\frac{3}{5}\, g_1^2 -3\, g_2^2 + 3\, Y_t^2+\frac{3}{2} \,\lambda_{STT}^2 + 3\, \lambda_{THH}^2\right.\notag\\
& \left.\quad\quad\quad\quad\quad+ 2 \,\kappa^2 + 4 \,\lambda_{SHH}^2 + 4\, \lambda_{SOO}^2 \right) \\
16\, \pi^2\, \frac{\text{d} \lambda_{STT}(t)}{\text{d}t} &= \frac{1}{2}\,\lambda_{STT}  \left( -16 \,g_2^2 + 4\, \lambda_{THH}^2 + 4\, \kappa^2 + 4 \,\lambda_{SHH}^2  + 7\, \lambda_{STT}^2 + 8\,\lambda_{SOO}^2 \right) \\
16\, \pi^2\, \frac{\text{d} \lambda_{THH}(t)}{\text{d}t} &= \lambda_{THH} \left( -\frac{3}{5} \,g_1^2 -7\, g_2^2 + 3 \,Y_t^2  +\lambda_{STT}^2 + 2\, \lambda_{SHH}^2 + 4 \,\lambda_{THH}^2 \right)\\
16\, \pi^2\, \frac{\text{d} \lambda_{SOO}(t)}{\text{d}t} &= \frac{\lambda_{SOO}}{2} \left( -24\, g_3^2 + 4\, \kappa^2 + 4\, \lambda_{SHH}^2 + 3\, \lambda_{STT}^2 + 12\,\lambda_{SOO}^2 \right)
\end{align}
where $t$ is $\log_{10} \Lambda/\gev$ and $\Lambda$ is the energy scale. The gauge couplings and the top Yukawa coupling were RGE evolved up to the 750 GeV mass scale using the SM RGEs. We then used the {\texttt{RunRGEs}} command within the \textsc{SARAH} \cite{Staub:2013tta,Staub:2012pb,Staub:2010jh} \textsc{Mathematica} package, matching to the SM values at the 750 GeV scale, and evolving up $10^{16} \gev$ or a Landau pole.

\section{Corrections to the Standard Model--like Higgs boson mass}
\label{app:HiggsMassCorrections}

The mass of the Standard Model--like Higgs is given by
\begin{eqnarray}
m_h^2 = m_Z^2\left[\cos^2 2\,\beta + \frac{2}{g_1^2+g_2^2}\left(\lambda_{SHH}^2+\frac{1}{2}\lambda_{THH}^2\right)\sin^2 2\,\beta\right] + \Delta_{m_h^2}^{(y_t)} + \Delta_{m_h^2}^{(\lambda)},
\end{eqnarray}
in the decoupling limit of large $m_{A^0}$. The loop contributions from the stops (with the stop mixing angle $\theta_t=0$) is given by~\cite{Martin:1997ns}
\begin{equation}
\Delta_{m_h^2}^{(y_t)}= \frac{3}{2\,\pi^2} \frac{m_t^4}{v^2} \left[ \log \frac{m_{\tilde{t}_1}\, m_{\tilde{t}_2}}{m^2_t} + \frac{\mu^2 \cot^2 \beta}{m_{\tilde{t}_1} m_{\tilde{t}_2}} \left(1-\frac{\mu^2 \cot^2\beta}{12 m_{\tilde{t}_1} m_{\tilde{t}_2}} \right) \right] .
\end{equation}
The $\Delta_{m_h^2}^{(\lambda)}$ term is given by~\cite{Benakli:2012cy}
\begin{equation}
\Delta_{m_h^2}^{(\lambda)} =  v^2 \left[ \lambda_1 \,\left(\cos^4 \beta +\, \sin^4 \beta\right) + 2 \left(\lambda_{SHH}^2 + \frac{1}{2}\,\lambda_{THH}^2 + \lambda_3^{\prime}\right) \cos^2\beta \,\sin^2\beta \right]
\end{equation}
and contains loop contributions from the singlet and triplet superfields, as well as the effects of integrating out the heavy $S$ and $T$ scalar fields. The values of $\lambda_1$ and $\lambda_3^{\prime}$ are 
\begin{align}
&128\,\pi^2 \,\lambda_1 = 4 \,\lambda^4_{SHH} \left(\llog{SI} + \llog{SR} \right)+ \lambda_{THH}^4 \left(\llog{TI} + \llog{TR} \right)\nonumber \\
& + 4\, \frac{\lambda_{SHH}^2 \lambda_{THH}^2}{\Big(\widetilde{m}^2_{SI} - \widetilde{m}^2_{TI}\Big)\Big(\widetilde{m}^2_{SR}-\widetilde{m}^2_{TR}\Big)} \bigg\{  -2\,\m{SI}\,\m{SR} + 2\, \m{SR}\,\m{TI} + 2\,\m{SI}\,\m{TR} - 2\,\m{TI}\,\m{TR} \nonumber \\
& ~  +\frac{\m{SI}}{\m{SR} - \m{TR}} \llog{SI} + \frac{\m{SR}}{\m{SI} - \m{TI}} \llog{SR} -\m{SR}\,\m{TI} \llog{TI} \nonumber \\
&~  +\m{TI}\,\m{TR}\,\llog{TI}  - \m{SI}\,\m{TR}\, \llog{TR} + \m{TI}\,\m{TR}\,\llog{TR} \bigg\} \nonumber \\
& +\frac{4}{\m{TI}-\m{TR}} \bigg\{ - \big(\m{TI}-\m{TR}\big)\big(-\lambda_{THH}^2+g_2^2\big)^2 \nonumber \\
&~ +\left( \frac{1}{2}\,\lambda_{THH}^4\big(3\,\m{TI}-\m{TR}\big) - 2\, g_2^2\,\lambda_{THH}^2\, \m{TI} +g_2^4\, \m{TI} \right)\llog{TI} \nonumber \\
&~ +\left( \frac{1}{2}\,\lambda_{THH}^4 \big(-3\,\m{TR}+\m{TI}\big)  + 2\, g_2^2\, \m{TR} \,\lambda_{THH}^2 - g_2^4\, \m{TR} \right) \llog{TR} \bigg\}
\end{align}
and 
\begin{align}
& 64\,\pi^2\, \lambda_3^{\prime} = 2\, \lambda_{SHH}^4 \left( \llog{SI} +\llog{SR} \right) + \frac{1}{2}\, \lambda_{THH}^4 \left( \llog{TI} + \llog{TR} \right) \nonumber \\
& - \frac{2 \,\lambda_{SHH}^2\,\lambda_{THH}^2}{\Big(\m{SI}-\m{TI}\Big)\Big(\m{SR}-\m{TR}\Big)} \bigg\{ -2\,\m{SI}\,\m{SR} +2\,\m{SR}\,\m{TI} + 2\, \m{SI}\, \m{TR} - 2\, \m{TI}\, \m{TR} \nonumber \\
&~ +\frac{\m{SI}}{\m{SR}-\m{TR}} \llog{SI} + \frac{\m{SR}}{\m{SI}-\m{TI}} \llog{SR} -\m{SR}\,\m{TI}\, \llog{TI}  \nonumber \\
&~ + \m{TI}\,\m{TR}\, \llog{TI}- \m{SI}\,\m{TR}\, \llog{TR} + \m{TI}\,\m{TR}\, \llog{TR} \bigg\} \nonumber\\
& + \frac{2}{\m{TI} - \m{TR}} \bigg\{ -\big(\m{TI} - \m{TR}\big)\big(-\lambda_{THH}^2 + g_2^2\big)^2  \nonumber \\[5pt]
&~ + \left( \frac{1}{2} \,\lambda_{THH}^4\,\big(3\,\m{TI}-\m{TR}\big)  - 2\, g_2^2 \,\lambda_{THH}^2\,\m{TI} + g_2^4\, \m{TI} \right) \llog{TI} \nonumber \\[5pt]
&~ + \left( \frac{1}{2} \,\lambda_{THH}^4\,\big(-3\,\m{TR}+\m{TI}\big)  + 2\, g_2^2 \,\lambda_{THH}^2\,\m{TR}  - g_2^4\,\m{TR} \right) \llog{TR} \bigg\}\,,
\end{align}
where $\widetilde{m}_{SR}$, $\widetilde{m}_{SI}$, $\widetilde{m}_{TR}$ and $\widetilde{m}_{TI}$ are the physical masses of the scalar and pseudoscalar of the singlet and triplet, respectively. For all the numerical results presented above, we have used the benchmark values
\begin{equation}
\widetilde{m}_{SR}=900\gev,\quad \widetilde{m}_{SI} = 750 \gev, ~\quad \widetilde{m}_{TR} = 1.4\tev, \quad \text{and } \quad \widetilde{m}_{TI} = 1.5\tev.
\end{equation}


\end{spacing}
\begin{spacing}{1.1}
\bibliography{DiracGauginosDiPhotons}

\providecommand{\href}[2]{#2}\begingroup\raggedright\begin{thebibliography}{100}

\bibitem{ATLAS-CONF-2015-081}
``{Search for resonances decaying to photon pairs in 3.2 fb$^{-1}$ of $pp$
  collisions at $\sqrt{s}$ = 13 TeV with the ATLAS detector},'' Tech. Rep.
  ATLAS-CONF-2015-081, CERN, Geneva, Dec, 2015.
\newblock \url{http://cds.cern.ch/record/2114853}.

\bibitem{ATLAS-CONF-2016-018}
``{Search for resonances in diphoton events with the ATLAS detector at
  $\sqrt{s}$ = 13 TeV},'' Tech. Rep. ATLAS-CONF-2016-018, CERN, Geneva, Mar,
  2016.
\newblock \url{http://cds.cern.ch/record/2141568}.

\bibitem{Aad:2014ioa}
{\bf ATLAS} Collaboration, G.~Aad {\em et al.}, ``{Search for Scalar Diphoton
  Resonances in the Mass Range $65-600$ GeV with the ATLAS Detector in $pp$
  Collision Data at $\sqrt{s}$ = 8 $TeV$},''
  \href{http://dx.doi.org/10.1103/PhysRevLett.113.171801}{{\em Phys. Rev.
  Lett.} {\bf 113} (2014) no.~17, 171801},
\href{http://arxiv.org/abs/1407.6583}{{\tt arXiv:1407.6583 [hep-ex]}}.

\bibitem{Aad:2015mna}
{\bf ATLAS} Collaboration, G.~Aad {\em et al.}, ``{Search for high-mass
  diphoton resonances in $pp$ collisions at $\sqrt{s}=8$ TeV with the ATLAS
  detector},'' \href{http://dx.doi.org/10.1103/PhysRevD.92.032004}{{\em Phys.
  Rev.} {\bf D92} (2015) no.~3, 032004},
\href{http://arxiv.org/abs/1504.05511}{{\tt arXiv:1504.05511 [hep-ex]}}.

\bibitem{CMS-PAS-EXO-15-004}
{\bf CMS Collaboration} Collaboration, ``{Search for new physics in high mass
  diphoton events in proton-proton collisions at $\sqrt{s} = 13$ TeV},'' Tech.
  Rep. CMS-PAS-EXO-15-004, CERN, Geneva, 2015.
\newblock \url{http://cds.cern.ch/record/2114808}.

\bibitem{CMS-PAS-EXO-16-018}
{\bf CMS Collaboration} Collaboration, ``{Search for new physics in high mass
  diphoton events in $3.3~\mathrm{fb}^{-1}$ of proton-proton collisions at
  $\sqrt{s}=13~\mathrm{TeV}$ and combined interpretation of searches at
  $8~\mathrm{TeV}$ and $13~\mathrm{TeV}$},'' Tech. Rep. CMS-PAS-EXO-16-018,
  CERN, Geneva, 2016.
\newblock \url{http://cds.cern.ch/record/2139899}.

\bibitem{Khachatryan:2015qba}
{\bf CMS} Collaboration, V.~Khachatryan {\em et al.}, ``{Search for diphoton
  resonances in the mass range from 150 to 850 GeV in pp collisions at
  $\sqrt{s} =$ 8 TeV},''
  \href{http://dx.doi.org/10.1016/j.physletb.2015.09.062}{{\em Phys. Lett.}
  {\bf B750} (2015)  494--519},
\href{http://arxiv.org/abs/1506.02301}{{\tt arXiv:1506.02301 [hep-ex]}}.

\bibitem{Chakrabortty:2015hff}
J.~Chakrabortty, A.~Choudhury, P.~Ghosh, S.~Mondal, and T.~Srivastava,
  ``{Di-photon resonance around 750 GeV: shedding light on the theory
  underneath},''
\href{http://arxiv.org/abs/1512.05767}{{\tt arXiv:1512.05767 [hep-ph]}}.

\bibitem{Buckley:2016mbr}
M.~R. Buckley, ``{Wide or narrow? The phenomenology of 750 GeV diphotons},''
  \href{http://dx.doi.org/10.1140/epjc/s10052-016-4201-y}{{\em Eur. Phys. J.}
  {\bf C76} (2016) no.~6, 345},
\href{http://arxiv.org/abs/1601.04751}{{\tt arXiv:1601.04751 [hep-ph]}}.

\bibitem{BuckleyBlog}
M.~R. Buckley, ``Diphotons: Moriond update.'' \url{
  http://www.physicsmatt.com/blog/2016/3/18/tdu3j18n86e01bd4a8l8u5flqdifce},
  March, 2016.

\bibitem{Franceschini:2015kwy}
R.~Franceschini, G.~F. Giudice, J.~F. Kamenik, M.~McCullough, A.~Pomarol,
  R.~Rattazzi, M.~Redi, F.~Riva, A.~Strumia, and R.~Torre, ``{What is the
  $\gamma \gamma$ resonance at 750 GeV?},''
  \href{http://dx.doi.org/10.1007/JHEP03(2016)144}{{\em JHEP} {\bf 03} (2016)
  144},
\href{http://arxiv.org/abs/1512.04933}{{\tt arXiv:1512.04933 [hep-ph]}}.

\bibitem{Knapen:2015dap}
S.~Knapen, T.~Melia, M.~Papucci, and K.~Zurek, ``{Rays of light from the
  LHC},'' \href{http://dx.doi.org/10.1103/PhysRevD.93.075020}{{\em Phys. Rev.}
  {\bf D93} (2016) no.~7, 075020},
\href{http://arxiv.org/abs/1512.04928}{{\tt arXiv:1512.04928 [hep-ph]}}.

\bibitem{Agrawal:2015dbf}
P.~Agrawal, J.~Fan, B.~Heidenreich, M.~Reece, and M.~Strassler, ``{Experimental
  Considerations Motivated by the Diphoton Excess at the LHC},''
  \href{http://dx.doi.org/10.1007/JHEP06(2016)082}{{\em JHEP} {\bf 06} (2016)
  082},
\href{http://arxiv.org/abs/1512.05775}{{\tt arXiv:1512.05775 [hep-ph]}}.

\bibitem{Gupta:2015zzs}
R.~S. Gupta, S.~Jager, Y.~Kats, G.~Perez, and E.~Stamou, ``{Interpreting a 750
  GeV Diphoton Resonance},''
\href{http://arxiv.org/abs/1512.05332}{{\tt arXiv:1512.05332 [hep-ph]}}.

\bibitem{Altmannshofer:2015xfo}
W.~Altmannshofer, J.~Galloway, S.~Gori, A.~L. Kagan, A.~Martin, and J.~Zupan,
  ``{750 GeV diphoton excess},''
  \href{http://dx.doi.org/10.1103/PhysRevD.93.095015}{{\em Phys. Rev.} {\bf
  D93} (2016) no.~9, 095015},
\href{http://arxiv.org/abs/1512.07616}{{\tt arXiv:1512.07616 [hep-ph]}}.

\bibitem{Craig:2015lra}
N.~Craig, P.~Draper, C.~Kilic, and S.~Thomas, ``{Shedding Light on Diphoton
  Resonances},'' \href{http://dx.doi.org/10.1103/PhysRevD.93.115023}{{\em Phys.
  Rev.} {\bf D93} (2016) no.~11, 115023},
\href{http://arxiv.org/abs/1512.07733}{{\tt arXiv:1512.07733 [hep-ph]}}.

\bibitem{Bernon:2016dow}
J.~Bernon, A.~Goudelis, S.~Kraml, K.~Mawatari, and D.~Sengupta,
  ``{Characterising the 750 GeV diphoton excess},''
  \href{http://dx.doi.org/10.1007/JHEP05(2016)128}{{\em JHEP} {\bf 05} (2016)
  128},
\href{http://arxiv.org/abs/1603.03421}{{\tt arXiv:1603.03421 [hep-ph]}}.

\bibitem{Kamenik:2016tuv}
J.~F. Kamenik, B.~R. Safdi, Y.~Soreq, and J.~Zupan, ``{Comments on the diphoton
  excess: critical reappraisal of effective field theory interpretations},''
  \href{http://dx.doi.org/10.1007/JHEP07(2016)042}{{\em JHEP} {\bf 07} (2016)
  042},
\href{http://arxiv.org/abs/1603.06566}{{\tt arXiv:1603.06566 [hep-ph]}}.

\bibitem{Franceschini:2016gxv}
R.~Franceschini, G.~F. Giudice, J.~F. Kamenik, M.~McCullough, F.~Riva,
  A.~Strumia, and R.~Torre, ``{Digamma, what next?},''
\href{http://arxiv.org/abs/1604.06446}{{\tt arXiv:1604.06446 [hep-ph]}}.

\bibitem{Backovic:2016xno}
M.~Backovic, ``{A Theory of Ambulance Chasing},''
\href{http://arxiv.org/abs/1603.01204}{{\tt arXiv:1603.01204
  [physics.soc-ph]}}.

\bibitem{Staub:2016dxq}
F.~Staub {\em et al.}, ``{Precision tools and models to narrow in on the 750
  GeV diphoton resonance},''
\href{http://arxiv.org/abs/1602.05581}{{\tt arXiv:1602.05581 [hep-ph]}}.

\bibitem{Petersson:2015mkr}
C.~Petersson and R.~Torre, ``{The 750 GeV diphoton excess from the goldstino
  superpartner},'' \href{http://dx.doi.org/10.1103/PhysRevLett.116.151804}{{\em
  Phys. Rev. Lett.} {\bf 116} (2016) no.~15, 151804},
\href{http://arxiv.org/abs/1512.05333}{{\tt arXiv:1512.05333 [hep-ph]}}.

\bibitem{Demidov:2015zqn}
S.~V. Demidov and D.~S. Gorbunov, ``{On the sgoldstino interpretation of the
  diphoton excess},'' \href{http://dx.doi.org/10.1134/S0021364016040044}{{\em
  JETP Lett.} {\bf 103} (2016) no.~4, 219--222},
\href{http://arxiv.org/abs/1512.05723}{{\tt arXiv:1512.05723 [hep-ph]}}.

\bibitem{Carpenter:2015ucu}
L.~M. Carpenter, R.~Colburn, and J.~Goodman, ``{Supersoft SUSY models and the
  750 GeV diphoton excess, beyond effective operators},''
  \href{http://dx.doi.org/10.1103/PhysRevD.94.015016}{{\em Phys. Rev.} {\bf
  D94} (2016) no.~1, 015016},
\href{http://arxiv.org/abs/1512.06107}{{\tt arXiv:1512.06107 [hep-ph]}}.

\bibitem{Feng:2015wil}
T.-F. Feng, X.-Q. Li, H.-B. Zhang, and S.-M. Zhao, ``{The LHC 750 GeV diphoton
  excess in supersymmetry with gauged baryon and lepton numbers},''
\href{http://arxiv.org/abs/1512.06696}{{\tt arXiv:1512.06696 [hep-ph]}}.

\bibitem{Ding:2015rxx}
R.~Ding, L.~Huang, T.~Li, and B.~Zhu, ``{Interpreting $750$ GeV Diphoton Excess
  with R-parity Violation Supersymmetry},''
\href{http://arxiv.org/abs/1512.06560}{{\tt arXiv:1512.06560 [hep-ph]}}.

\bibitem{Wang:2015kuj}
F.~Wang, L.~Wu, J.~M. Yang, and M.~Zhang, ``{750 GeV diphoton resonance, 125
  GeV Higgs and muon g − 2 anomaly in deflected anomaly mediation SUSY
  breaking scenarios},''
  \href{http://dx.doi.org/10.1016/j.physletb.2016.05.071}{{\em Phys. Lett.}
  {\bf B759} (2016)  191--199},
\href{http://arxiv.org/abs/1512.06715}{{\tt arXiv:1512.06715 [hep-ph]}}.

\bibitem{Chakraborty:2015gyj}
S.~Chakraborty, A.~Chakraborty, and S.~Raychaudhuri, ``{Diphoton resonance at
  750 GeV in the broken MRSSM},''
\href{http://arxiv.org/abs/1512.07527}{{\tt arXiv:1512.07527 [hep-ph]}}.

\bibitem{Allanach:2015ixl}
B.~C. Allanach, P.~S.~B. Dev, S.~A. Renner, and K.~Sakurai, ``{750 GeV diphoton
  excess explained by a resonant sneutrino in R-parity violating
  supersymmetry},'' \href{http://dx.doi.org/10.1103/PhysRevD.93.115022}{{\em
  Phys. Rev.} {\bf D93} (2016) no.~11, 115022},
\href{http://arxiv.org/abs/1512.07645}{{\tt arXiv:1512.07645 [hep-ph]}}.

\bibitem{Casas:2015blx}
J.~A. Casas, J.~R. Espinosa, and J.~M. Moreno, ``{The 750 GeV Diphoton Excess
  as a First Light on Supersymmetry Breaking},''
  \href{http://dx.doi.org/10.1016/j.physletb.2016.05.070}{{\em Phys. Lett.}
  {\bf B759} (2016)  159--165},
\href{http://arxiv.org/abs/1512.07895}{{\tt arXiv:1512.07895 [hep-ph]}}.

\bibitem{Hall:2015xds}
L.~J. Hall, K.~Harigaya, and Y.~Nomura, ``{750 GeV Diphotons: Implications for
  Supersymmetric Unification},''
  \href{http://dx.doi.org/10.1007/JHEP03(2016)017}{{\em JHEP} {\bf 03} (2016)
  017},
\href{http://arxiv.org/abs/1512.07904}{{\tt arXiv:1512.07904 [hep-ph]}}.

\bibitem{Hall:2016swn}
L.~J. Hall, K.~Harigaya, and Y.~Nomura, ``{750 GeV Diphotons: Implications for
  Supersymmetric Unification II},''
\href{http://arxiv.org/abs/1605.03585}{{\tt arXiv:1605.03585 [hep-ph]}}.

\bibitem{Wang:2015omi}
F.~Wang, W.~Wang, L.~Wu, J.~M. Yang, and M.~Zhang, ``{Interpreting 750 GeV
  diphoton resonance as degenerate Higgs bosons in NMSSM with vector-like
  particles},''
\href{http://arxiv.org/abs/1512.08434}{{\tt arXiv:1512.08434 [hep-ph]}}.

\bibitem{Tang:2015eko}
Y.-L. Tang and S.-h. Zhu, ``{NMSSM extended with vector-like particles and the
  diphoton excess on the LHC},''
\href{http://arxiv.org/abs/1512.08323}{{\tt arXiv:1512.08323 [hep-ph]}}.

\bibitem{Chao:2016mtn}
W.~Chao, ``{The Diphoton Excess from an Exceptional Supersymmetric Standard
  Model},''
\href{http://arxiv.org/abs/1601.00633}{{\tt arXiv:1601.00633 [hep-ph]}}.

\bibitem{Dutta:2016jqn}
B.~Dutta, Y.~Gao, T.~Ghosh, I.~Gogoladze, T.~Li, Q.~Shafi, and J.~W. Walker,
  ``{Diphoton Excess in Consistent Supersymmetric SU(5) Models with Vector-like
  Particles},''
\href{http://arxiv.org/abs/1601.00866}{{\tt arXiv:1601.00866 [hep-ph]}}.

\bibitem{King:2016wep}
S.~F. King and R.~Nevzorov, ``{750 GeV Diphoton Resonance from Singlets in an
  Exceptional Supersymmetric Standard Model},''
  \href{http://dx.doi.org/10.1007/JHEP03(2016)139}{{\em JHEP} {\bf 03} (2016)
  139},
\href{http://arxiv.org/abs/1601.07242}{{\tt arXiv:1601.07242 [hep-ph]}}.

\bibitem{Ding:2016udc}
R.~Ding, Y.~Fan, L.~Huang, C.~Li, T.~Li, S.~Raza, and B.~Zhu, ``{Systematic
  Study of Diphoton Resonance at 750 GeV from Sgoldstino},''
\href{http://arxiv.org/abs/1602.00977}{{\tt arXiv:1602.00977 [hep-ph]}}.

\bibitem{Ellwanger:2016qax}
U.~Ellwanger and C.~Hugonie, ``{A 750 GeV Diphoton Signal from a Very Light
  Pseudoscalar in the NMSSM},''
  \href{http://dx.doi.org/10.1007/JHEP05(2016)114}{{\em JHEP} {\bf 05} (2016)
  114},
\href{http://arxiv.org/abs/1602.03344}{{\tt arXiv:1602.03344 [hep-ph]}}.

\bibitem{Han:2016fli}
C.~Han, T.~T. Yanagida, and N.~Yokozaki, ``{Implications of the 750 GeV
  Diphoton Excess in Gaugino Mediation},''
  \href{http://dx.doi.org/10.1103/PhysRevD.93.055025}{{\em Phys. Rev.} {\bf
  D93} (2016) no.~5, 055025},
\href{http://arxiv.org/abs/1602.04204}{{\tt arXiv:1602.04204 [hep-ph]}}.

\bibitem{Barbieri:2016cnt}
R.~Barbieri, D.~Buttazzo, L.~J. Hall, and D.~Marzocca, ``{Higgs mass and
  unified gauge coupling in the NMSSM with Vector Matter},''
  \href{http://dx.doi.org/10.1007/JHEP07(2016)067}{{\em JHEP} {\bf 07} (2016)
  067},
\href{http://arxiv.org/abs/1603.00718}{{\tt arXiv:1603.00718 [hep-ph]}}.

\bibitem{Badziak:2016cfd}
M.~Badziak, M.~Olechowski, S.~Pokorski, and K.~Sakurai, ``{Interpreting 750 GeV
  Diphoton Excess in Plain NMSSM},''
  \href{http://dx.doi.org/10.1016/j.physletb.2016.06.057}{{\em Phys. Lett.}
  {\bf B760} (2016)  228--235},
\href{http://arxiv.org/abs/1603.02203}{{\tt arXiv:1603.02203 [hep-ph]}}.

\bibitem{Baratella:2016daa}
P.~Baratella, J.~Elias-Miro, J.~Penedo, and A.~Romanino, ``{A closer look to
  the sgoldstino interpretation of the diphoton excess},''
  \href{http://dx.doi.org/10.1007/JHEP06(2016)086}{{\em JHEP} {\bf 06} (2016)
  086},
\href{http://arxiv.org/abs/1603.05682}{{\tt arXiv:1603.05682 [hep-ph]}}.

\bibitem{Nilles:2016bjl}
H.~P. Nilles and M.~W. Winkler, ``{750 GeV Diphotons and Supersymmetric Grand
  Unification},'' \href{http://dx.doi.org/10.1007/JHEP05(2016)182}{{\em JHEP}
  {\bf 05} (2016)  182},
\href{http://arxiv.org/abs/1604.03598}{{\tt arXiv:1604.03598 [hep-ph]}}.

\bibitem{Gu:2015lxj}
J.~Gu and Z.~Liu, ``{Physics implications of the diphoton excess from the
  perspective of renormalization group flow},''
  \href{http://dx.doi.org/10.1103/PhysRevD.93.075006}{{\em Phys. Rev.} {\bf
  D93} (2016) no.~7, 075006},
\href{http://arxiv.org/abs/1512.07624}{{\tt arXiv:1512.07624 [hep-ph]}}.

\bibitem{Bae:2016xni}
K.~J. Bae, M.~Endo, K.~Hamaguchi, and T.~Moroi, ``{Diphoton Excess and Running
  Couplings},'' \href{http://dx.doi.org/10.1016/j.physletb.2016.04.031}{{\em
  Phys. Lett.} {\bf B757} (2016)  493--500},
\href{http://arxiv.org/abs/1602.03653}{{\tt arXiv:1602.03653 [hep-ph]}}.

\bibitem{Hamada:2016vwk}
Y.~Hamada, H.~Kawai, K.~Kawana, and K.~Tsumura, ``{Models of the LHC diphoton
  excesses valid up to the Planck scale},''
  \href{http://dx.doi.org/10.1103/PhysRevD.94.014007}{{\em Phys. Rev.} {\bf
  D94} (2016) no.~1, 014007},
\href{http://arxiv.org/abs/1602.04170}{{\tt arXiv:1602.04170 [hep-ph]}}.

\bibitem{Choudhury:2016jbc}
D.~Choudhury and K.~Ghosh, ``{The LHC Diphoton excess at 750 GeV in the
  framework of the Constrained Minimal Supersymmetric Standard Model},''
\href{http://arxiv.org/abs/1605.00013}{{\tt arXiv:1605.00013 [hep-ph]}}.

\bibitem{Djouadi:2016oey}
A.~Djouadi and A.~Pilaftsis, ``{The 750 GeV Diphoton Resonance in the MSSM},''
\href{http://arxiv.org/abs/1605.01040}{{\tt arXiv:1605.01040 [hep-ph]}}.

\bibitem{Bardhan:2016rsb}
D.~Bardhan, P.~Byakti, D.~Ghosh, and T.~Sharma, ``{The 750 GeV diphoton
  resonance as an sgoldstino: a reappraisal},''
  \href{http://dx.doi.org/10.1007/JHEP06(2016)129}{{\em JHEP} {\bf 06} (2016)
  129},
\href{http://arxiv.org/abs/1603.05251}{{\tt arXiv:1603.05251 [hep-ph]}}.

\bibitem{Bharucha:2016jyr}
A.~Bharucha, A.~Djouadi, and A.~Goudelis, ``{Threshold enhancement of diphoton
  resonances},''
\href{http://arxiv.org/abs/1603.04464}{{\tt arXiv:1603.04464 [hep-ph]}}.

\bibitem{Fox:2002bu}
P.~J. Fox, A.~E. Nelson, and N.~Weiner, ``{Dirac gaugino masses and supersoft
  supersymmetry breaking},''
  \href{http://dx.doi.org/10.1088/1126-6708/2002/08/035}{{\em JHEP} {\bf 08}
  (2002)  035},
\href{http://arxiv.org/abs/hep-ph/0206096}{{\tt arXiv:hep-ph/0206096
  [hep-ph]}}.

\bibitem{Itoyama:2011zi}
H.~Itoyama and N.~Maru, ``{D-term Dynamical Supersymmetry Breaking Generating
  Split N=2 Gaugino Masses of Mixed Majorana-Dirac Type},''
  \href{http://dx.doi.org/10.1142/S0217751X1250159X}{{\em Int. J. Mod. Phys.}
  {\bf A27} (2012)  1250159},
\href{http://arxiv.org/abs/1109.2276}{{\tt arXiv:1109.2276 [hep-ph]}}.

\bibitem{Itoyama:2013sn}
H.~Itoyama and N.~Maru, ``{D-term Triggered Dynamical Supersymmetry
  Breaking},'' \href{http://dx.doi.org/10.1103/PhysRevD.88.025012}{{\em Phys.
  Rev.} {\bf D88} (2013) no.~2, 025012},
\href{http://arxiv.org/abs/1301.7548}{{\tt arXiv:1301.7548 [hep-ph]}}.

\bibitem{Itoyama:2013vxa}
H.~Itoyama and N.~Maru, ``{126 GeV Higgs Boson Associated with D-term Triggered
  Dynamical Supersymmetry Breaking},''
  \href{http://dx.doi.org/10.3390/sym7010193}{{\em Symmetry} {\bf 7} (2015)
  no.~1, 193--205},
\href{http://arxiv.org/abs/1312.4157}{{\tt arXiv:1312.4157 [hep-ph]}}.

\bibitem{Alves:2015kia}
D.~S.~M. Alves, J.~Galloway, M.~McCullough, and N.~Weiner, ``{Goldstone
  Gauginos},'' \href{http://dx.doi.org/10.1103/PhysRevLett.115.161801}{{\em
  Phys. Rev. Lett.} {\bf 115} (2015) no.~16, 161801},
\href{http://arxiv.org/abs/1502.03819}{{\tt arXiv:1502.03819 [hep-ph]}}.

\bibitem{Martin:2015eca}
S.~P. Martin, ``{Nonstandard supersymmetry breaking and Dirac gaugino masses
  without supersoftness},''
  \href{http://dx.doi.org/10.1103/PhysRevD.92.035004}{{\em Phys. Rev.} {\bf
  D92} (2015) no.~3, 035004},
\href{http://arxiv.org/abs/1506.02105}{{\tt arXiv:1506.02105 [hep-ph]}}.

\bibitem{Kribs:2012gx}
G.~D. Kribs and A.~Martin, ``{Supersoft Supersymmetry is Super-Safe},''
  \href{http://dx.doi.org/10.1103/PhysRevD.85.115014}{{\em Phys. Rev.} {\bf
  D85} (2012)  115014},
\href{http://arxiv.org/abs/1203.4821}{{\tt arXiv:1203.4821 [hep-ph]}}.

\bibitem{Kribs:2013eua}
G.~D. Kribs and N.~Raj, ``{Mixed Gauginos Sending Mixed Messages to the LHC},''
  \href{http://dx.doi.org/10.1103/PhysRevD.89.055011}{{\em Phys. Rev.} {\bf
  D89} (2014) no.~5, 055011},
\href{http://arxiv.org/abs/1307.7197}{{\tt arXiv:1307.7197 [hep-ph]}}.

\bibitem{Cohen:2013xda}
T.~Cohen, T.~Golling, M.~Hance, A.~Henrichs, K.~Howe, J.~Loyal, S.~Padhi, and
  J.~G. Wacker, ``{SUSY Simplified Models at 14, 33, and 100 TeV Proton
  Colliders},'' \href{http://dx.doi.org/10.1007/JHEP04(2014)117}{{\em JHEP}
  {\bf 04} (2014)  117},
\href{http://arxiv.org/abs/1311.6480}{{\tt arXiv:1311.6480 [hep-ph]}}.

\bibitem{Cohen:2014hxa}
T.~Cohen, R.~T. D'Agnolo, M.~Hance, H.~K. Lou, and J.~G. Wacker, ``{Boosting
  Stop Searches with a 100 TeV Proton Collider},''
  \href{http://dx.doi.org/10.1007/JHEP11(2014)021}{{\em JHEP} {\bf 11} (2014)
  021},
\href{http://arxiv.org/abs/1406.4512}{{\tt arXiv:1406.4512 [hep-ph]}}.

\bibitem{Gori:2014oua}
S.~Gori, S.~Jung, L.-T. Wang, and J.~D. Wells, ``{Prospects for Electroweakino
  Discovery at a 100 TeV Hadron Collider},''
  \href{http://dx.doi.org/10.1007/JHEP12(2014)108}{{\em JHEP} {\bf 12} (2014)
  108},
\href{http://arxiv.org/abs/1410.6287}{{\tt arXiv:1410.6287 [hep-ph]}}.

\bibitem{Arkani-Hamed:2015vfh}
N.~Arkani-Hamed, T.~Han, M.~Mangano, and L.-T. Wang, ``{Physics Opportunities
  of a 100 TeV Proton-Proton Collider},''
\href{http://arxiv.org/abs/1511.06495}{{\tt arXiv:1511.06495 [hep-ph]}}.

\bibitem{Bramante:2015una}
J.~Bramante, N.~Desai, P.~Fox, A.~Martin, B.~Ostdiek, and T.~Plehn, ``{Towards
  the Final Word on Neutralino Dark Matter},''
  \href{http://dx.doi.org/10.1103/PhysRevD.93.063525}{{\em Phys. Rev.} {\bf
  D93} (2016) no.~6, 063525},
\href{http://arxiv.org/abs/1510.03460}{{\tt arXiv:1510.03460 [hep-ph]}}.

\bibitem{Bramante:2014tba}
J.~Bramante, P.~J. Fox, A.~Martin, B.~Ostdiek, T.~Plehn, T.~Schell, and
  M.~Takeuchi, ``{Relic neutralino surface at a 100 TeV collider},''
  \href{http://dx.doi.org/10.1103/PhysRevD.91.054015}{{\em Phys. Rev.} {\bf
  D91} (2015)  054015},
\href{http://arxiv.org/abs/1412.4789}{{\tt arXiv:1412.4789 [hep-ph]}}.

\bibitem{Kribs:2007ac}
G.~D. Kribs, E.~Poppitz, and N.~Weiner, ``{Flavor in supersymmetry with an
  extended R-symmetry},''
  \href{http://dx.doi.org/10.1103/PhysRevD.78.055010}{{\em Phys. Rev.} {\bf
  D78} (2008)  055010},
\href{http://arxiv.org/abs/0712.2039}{{\tt arXiv:0712.2039 [hep-ph]}}.

\bibitem{Djouadi:2005gi}
A.~Djouadi, ``{The Anatomy of electro-weak symmetry breaking. I: The Higgs
  boson in the standard model},''
  \href{http://dx.doi.org/10.1016/j.physrep.2007.10.004}{{\em Phys. Rept.} {\bf
  457} (2008)  1--216},
\href{http://arxiv.org/abs/hep-ph/0503172}{{\tt arXiv:hep-ph/0503172
  [hep-ph]}}.

\bibitem{Low:2015qep}
M.~Low, A.~Tesi, and L.-T. Wang, ``{A pseudoscalar decaying to photon pairs in
  the early LHC Run 2 data},''
  \href{http://dx.doi.org/10.1007/JHEP03(2016)108}{{\em JHEP} {\bf 03} (2016)
  108},
\href{http://arxiv.org/abs/1512.05328}{{\tt arXiv:1512.05328 [hep-ph]}}.

\bibitem{Berthier:2015vbb}
L.~Berthier, J.~M. Cline, W.~Shepherd, and M.~Trott, ``{Effective
  interpretations of a diphoton excess},''
  \href{http://dx.doi.org/10.1007/JHEP04(2016)084}{{\em JHEP} {\bf 04} (2016)
  084},
\href{http://arxiv.org/abs/1512.06799}{{\tt arXiv:1512.06799 [hep-ph]}}.

\bibitem{Bai:2016czm}
Y.~Bai, V.~Barger, and J.~Berger, ``{Constraints on color-octet companions of a
  750 GeV heavy pion from dijet and photon plus jet resonance searches},''
  \href{http://dx.doi.org/10.1103/PhysRevD.94.011701}{{\em Phys. Rev.} {\bf
  D94} (2016) no.~1, 011701},
\href{http://arxiv.org/abs/1604.07835}{{\tt arXiv:1604.07835 [hep-ph]}}.

\bibitem{Harigaya:2016eol}
K.~Harigaya and Y.~Nomura, ``{Hidden Pion Varieties in Composite Models for
  Diphoton Resonances},''
\href{http://arxiv.org/abs/1603.05774}{{\tt arXiv:1603.05774 [hep-ph]}}.

\bibitem{Draper:2016fsr}
P.~Draper and D.~McKeen, ``{Diphotons, New Vacuum Angles, and Strong CP},''
  \href{http://dx.doi.org/10.1007/JHEP04(2016)127}{{\em JHEP} {\bf 04} (2016)
  127},
\href{http://arxiv.org/abs/1602.03604}{{\tt arXiv:1602.03604 [hep-ph]}}.

\bibitem{Howe:2016mfq}
K.~Howe, S.~Knapen, and D.~J. Robinson, ``{Diphotons from an Electroweak
  Triplet-Singlet},''
\href{http://arxiv.org/abs/1603.08932}{{\tt arXiv:1603.08932 [hep-ph]}}.

\bibitem{Aad:2014vma}
{\bf ATLAS} Collaboration, G.~Aad {\em et al.}, ``{Search for direct production
  of charginos, neutralinos and sleptons in final states with two leptons and
  missing transverse momentum in $pp$ collisions at $\sqrt{s} =$ 8 TeV with the
  ATLAS detector},'' \href{http://dx.doi.org/10.1007/JHEP05(2014)071}{{\em
  JHEP} {\bf 05} (2014)  071},
\href{http://arxiv.org/abs/1403.5294}{{\tt arXiv:1403.5294 [hep-ex]}}.

\bibitem{Aad:2015jqa}
{\bf ATLAS} Collaboration, G.~Aad {\em et al.}, ``{Search for direct pair
  production of a chargino and a neutralino decaying to the 125 GeV Higgs boson
  in $\sqrt{s} = 8$ TeV ${pp}$ collisions with the ATLAS detector},''
  \href{http://dx.doi.org/10.1140/epjc/s10052-015-3408-7}{{\em Eur. Phys. J.}
  {\bf C75} (2015) no.~5, 208},
\href{http://arxiv.org/abs/1501.07110}{{\tt arXiv:1501.07110 [hep-ex]}}.

\bibitem{Khachatryan:2014qwa}
{\bf CMS} Collaboration, V.~Khachatryan {\em et al.}, ``{Searches for
  electroweak production of charginos, neutralinos, and sleptons decaying to
  leptons and W, Z, and Higgs bosons in pp collisions at 8 TeV},''
  \href{http://dx.doi.org/10.1140/epjc/s10052-014-3036-7}{{\em Eur. Phys. J.}
  {\bf C74} (2014) no.~9, 3036},
\href{http://arxiv.org/abs/1405.7570}{{\tt arXiv:1405.7570 [hep-ex]}}.

\bibitem{Khachatryan:2014mma}
{\bf CMS} Collaboration, V.~Khachatryan {\em et al.}, ``{Searches for
  electroweak neutralino and chargino production in channels with Higgs, Z, and
  W bosons in pp collisions at 8 TeV},''
  \href{http://dx.doi.org/10.1103/PhysRevD.90.092007}{{\em Phys. Rev.} {\bf
  D90} (2014) no.~9, 092007},
\href{http://arxiv.org/abs/1409.3168}{{\tt arXiv:1409.3168 [hep-ex]}}.

\bibitem{Arina:2014xya}
C.~Arina, V.~Martin-Lozano, and G.~Nardini, ``{Dark matter versus $h \to
  \gamma\gamma$ and $h \to \gamma Z$ with supersymmetric triplets},''
  \href{http://dx.doi.org/10.1007/JHEP08(2014)015}{{\em JHEP} {\bf 08} (2014)
  015},
\href{http://arxiv.org/abs/1403.6434}{{\tt arXiv:1403.6434 [hep-ph]}}.

\bibitem{Basak:2013eba}
T.~Basak and S.~Mohanty, ``{130 GeV gamma ray line and enhanced Higgs di-photon
  rate from Triplet-Singlet extended MSSM},''
  \href{http://dx.doi.org/10.1007/JHEP08(2013)020}{{\em JHEP} {\bf 08} (2013)
  020},
\href{http://arxiv.org/abs/1304.6856}{{\tt arXiv:1304.6856 [hep-ph]}}.

\bibitem{deBlas:2013epa}
J.~de~Blas, A.~Delgado, B.~Ostdiek, and M.~Quirós, ``{Indirect effects of
  supersymmetric triplets in stop decays},''
  \href{http://dx.doi.org/10.1007/JHEP01(2014)177}{{\em JHEP} {\bf 01} (2014)
  177},
\href{http://arxiv.org/abs/1311.3654}{{\tt arXiv:1311.3654 [hep-ph]}}.

\bibitem{Delgado:2012sm}
A.~Delgado, G.~Nardini, and M.~Quiros, ``{Large diphoton Higgs rates from
  supersymmetric triplets},''
  \href{http://dx.doi.org/10.1103/PhysRevD.86.115010}{{\em Phys. Rev.} {\bf
  D86} (2012)  115010},
\href{http://arxiv.org/abs/1207.6596}{{\tt arXiv:1207.6596 [hep-ph]}}.

\bibitem{DiChiara:2008rg}
S.~Di~Chiara and K.~Hsieh, ``{Triplet Extended Supersymmetric Standard
  Model},'' \href{http://dx.doi.org/10.1103/PhysRevD.78.055016}{{\em Phys.
  Rev.} {\bf D78} (2008)  055016},
\href{http://arxiv.org/abs/0805.2623}{{\tt arXiv:0805.2623 [hep-ph]}}.

\bibitem{Staub:2010jh}
F.~Staub, ``{Automatic Calculation of supersymmetric Renormalization Group
  Equations and Self Energies},''
  \href{http://dx.doi.org/10.1016/j.cpc.2010.11.030}{{\em Comput. Phys.
  Commun.} {\bf 182} (2011)  808--833},
\href{http://arxiv.org/abs/1002.0840}{{\tt arXiv:1002.0840 [hep-ph]}}.

\bibitem{Staub:2012pb}
F.~Staub, ``{SARAH 3.2: Dirac Gauginos, UFO output, and more},''
  \href{http://dx.doi.org/10.1016/j.cpc.2013.02.019}{{\em Comput. Phys.
  Commun.} {\bf 184} (2013)  1792--1809},
\href{http://arxiv.org/abs/1207.0906}{{\tt arXiv:1207.0906 [hep-ph]}}.

\bibitem{Staub:2013tta}
F.~Staub, ``{SARAH 4 : A tool for (not only SUSY) model builders},''
  \href{http://dx.doi.org/10.1016/j.cpc.2014.02.018}{{\em Comput. Phys.
  Commun.} {\bf 185} (2014)  1773--1790},
\href{http://arxiv.org/abs/1309.7223}{{\tt arXiv:1309.7223 [hep-ph]}}.

\bibitem{Martin:1997ns}
S.~P. Martin, ``{A Supersymmetry primer},''
  \href{http://arxiv.org/abs/hep-ph/9709356}{{\tt arXiv:hep-ph/9709356
  [hep-ph]}}.
[Adv. Ser. Direct. High Energy Phys.18,1(1998)].

\bibitem{Ellwanger:2005fh}
U.~Ellwanger and C.~Hugonie, ``{Yukawa induced radiative corrections to the
  lightest Higgs boson mass in the NMSSM},''
  \href{http://dx.doi.org/10.1016/j.physletb.2005.07.039}{{\em Phys. Lett.}
  {\bf B623} (2005)  93--103},
\href{http://arxiv.org/abs/hep-ph/0504269}{{\tt arXiv:hep-ph/0504269
  [hep-ph]}}.

\bibitem{Ender:2011qh}
K.~Ender, T.~Graf, M.~Muhlleitner, and H.~Rzehak, ``{Analysis of the NMSSM
  Higgs Boson Masses at One-Loop Level},''
  \href{http://dx.doi.org/10.1103/PhysRevD.85.075024}{{\em Phys. Rev.} {\bf
  D85} (2012)  075024},
\href{http://arxiv.org/abs/1111.4952}{{\tt arXiv:1111.4952 [hep-ph]}}.

\bibitem{Benakli:2011kz}
K.~Benakli, M.~D. Goodsell, and A.-K. Maier, ``{Generating mu and Bmu in models
  with Dirac Gauginos},''
  \href{http://dx.doi.org/10.1016/j.nuclphysb.2011.06.001}{{\em Nucl. Phys.}
  {\bf B851} (2011)  445--461},
\href{http://arxiv.org/abs/1104.2695}{{\tt arXiv:1104.2695 [hep-ph]}}.

\bibitem{Benakli:2012cy}
K.~Benakli, M.~D. Goodsell, and F.~Staub, ``{Dirac Gauginos and the 125 GeV
  Higgs},'' \href{http://dx.doi.org/10.1007/JHEP06(2013)073}{{\em JHEP} {\bf
  06} (2013)  073},
\href{http://arxiv.org/abs/1211.0552}{{\tt arXiv:1211.0552 [hep-ph]}}.

\bibitem{Aad:2015pfx}
{\bf ATLAS} Collaboration, G.~Aad {\em et al.}, ``{ATLAS Run 1 searches for
  direct pair production of third-generation squarks at the Large Hadron
  Collider},'' \href{http://dx.doi.org/10.1140/epjc/s10052-015-3726-9,
  10.1140/epjc/s10052-016-3935-x}{{\em Eur. Phys. J.} {\bf C75} (2015) no.~10,
  510}, \href{http://arxiv.org/abs/1506.08616}{{\tt arXiv:1506.08616
  [hep-ex]}}.
[Erratum: Eur. Phys. J.C76,no.3,153(2016)].

\bibitem{Aad:2014bva}
{\bf ATLAS} Collaboration, G.~Aad {\em et al.}, ``{Search for direct pair
  production of the top squark in all-hadronic final states in proton-proton
  collisions at $\sqrt{s}=8$ TeV with the ATLAS detector},''
  \href{http://dx.doi.org/10.1007/JHEP09(2014)015}{{\em JHEP} {\bf 09} (2014)
  015},
\href{http://arxiv.org/abs/1406.1122}{{\tt arXiv:1406.1122 [hep-ex]}}.

\bibitem{Aad:2014kra}
{\bf ATLAS} Collaboration, G.~Aad {\em et al.}, ``{Search for top squark pair
  production in final states with one isolated lepton, jets, and missing
  transverse momentum in $\sqrt s =$8 TeV $pp$ collisions with the ATLAS
  detector},'' \href{http://dx.doi.org/10.1007/JHEP11(2014)118}{{\em JHEP} {\bf
  11} (2014)  118},
\href{http://arxiv.org/abs/1407.0583}{{\tt arXiv:1407.0583 [hep-ex]}}.

\bibitem{Aad:2014qaa}
{\bf ATLAS} Collaboration, G.~Aad {\em et al.}, ``{Search for direct top-squark
  pair production in final states with two leptons in pp collisions at
  $\sqrt{s} =$ 8TeV with the ATLAS detector},''
  \href{http://dx.doi.org/10.1007/JHEP06(2014)124}{{\em JHEP} {\bf 06} (2014)
  124},
\href{http://arxiv.org/abs/1403.4853}{{\tt arXiv:1403.4853 [hep-ex]}}.

\bibitem{Aad:2014nra}
{\bf ATLAS} Collaboration, G.~Aad {\em et al.}, ``{Search for pair-produced
  third-generation squarks decaying via charm quarks or in compressed
  supersymmetric scenarios in $pp$ collisions at $\sqrt{s}=8~$TeV with the
  ATLAS detector},'' \href{http://dx.doi.org/10.1103/PhysRevD.90.052008}{{\em
  Phys. Rev.} {\bf D90} (2014) no.~5, 052008},
\href{http://arxiv.org/abs/1407.0608}{{\tt arXiv:1407.0608 [hep-ex]}}.

\bibitem{Aad:2014mha}
{\bf ATLAS} Collaboration, G.~Aad {\em et al.}, ``{Search for direct top squark
  pair production in events with a Z boson, b-jets and missing transverse
  momentum in sqrt(s)=8 TeV pp collisions with the ATLAS detector},''
  \href{http://dx.doi.org/10.1140/epjc/s10052-014-2883-6}{{\em Eur. Phys. J.}
  {\bf C74} (2014) no.~6, 2883},
\href{http://arxiv.org/abs/1403.5222}{{\tt arXiv:1403.5222 [hep-ex]}}.

\bibitem{ATLAS:2014fka}
{\bf ATLAS} Collaboration, G.~Aad {\em et al.}, ``{Searches for heavy
  long-lived charged particles with the ATLAS detector in proton-proton
  collisions at $ \sqrt{s}=8 $ TeV},''
  \href{http://dx.doi.org/10.1007/JHEP01(2015)068}{{\em JHEP} {\bf 01} (2015)
  068},
\href{http://arxiv.org/abs/1411.6795}{{\tt arXiv:1411.6795 [hep-ex]}}.

\bibitem{Aad:2013gva}
{\bf ATLAS} Collaboration, G.~Aad {\em et al.}, ``{Search for long-lived
  stopped R-hadrons decaying out-of-time with pp collisions using the ATLAS
  detector},'' \href{http://dx.doi.org/10.1103/PhysRevD.88.112003}{{\em Phys.
  Rev.} {\bf D88} (2013) no.~11, 112003},
\href{http://arxiv.org/abs/1310.6584}{{\tt arXiv:1310.6584 [hep-ex]}}.

\bibitem{Khachatryan:2016xvy}
{\bf CMS} Collaboration, V.~Khachatryan {\em et al.}, ``{Search for new physics
  with the MT2 variable in all-jets final states produced in pp collisions at
  sqrt(s) = 13 TeV},''
\href{http://arxiv.org/abs/1603.04053}{{\tt arXiv:1603.04053 [hep-ex]}}.

\bibitem{Khachatryan:2016kdk}
{\bf CMS} Collaboration, V.~Khachatryan {\em et al.}, ``{Search for
  supersymmetry in the multijet and missing transverse momentum final state in
  pp collisions at 13 TeV},''
  \href{http://dx.doi.org/10.1016/j.physletb.2016.05.002}{{\em Phys. Lett.}
  {\bf B758} (2016)  152--180},
\href{http://arxiv.org/abs/1602.06581}{{\tt arXiv:1602.06581 [hep-ex]}}.

\bibitem{CMS:2016qtx}
{\bf CMS} Collaboration, C.~Collaboration,
``{Search for direct top squark pair production in the single lepton final
  state at $\sqrt{s}=13~\mathrm{TeV}$},''.

\bibitem{Khachatryan:2016zcu}
{\bf CMS} Collaboration, V.~Khachatryan {\em et al.}, ``{Search for
  supersymmetry in pp collisions at sqrt(s) = 8 TeV in final states with
  boosted W bosons and b jets using razor variables},''
  \href{http://dx.doi.org/10.1103/PhysRevD.93.092009}{{\em Phys. Rev.} {\bf
  D93} (2016) no.~9, 092009},
\href{http://arxiv.org/abs/1602.02917}{{\tt arXiv:1602.02917 [hep-ex]}}.

\bibitem{CMS:2014dpa}
{\bf CMS} Collaboration, V.~Khachatryan {\em et al.}, ``{Searches for
  supersymmetry based on events with b jets and four W bosons in pp collisions
  at 8 TeV},'' \href{http://dx.doi.org/10.1016/j.physletb.2015.04.002}{{\em
  Phys. Lett.} {\bf B745} (2015)  5--28},
\href{http://arxiv.org/abs/1412.4109}{{\tt arXiv:1412.4109 [hep-ex]}}.

\bibitem{Khachatryan:2014doa}
{\bf CMS} Collaboration, V.~Khachatryan {\em et al.}, ``{Search for top-squark
  pairs decaying into Higgs or Z bosons in pp collisions at $\sqrt{s}$=8
  TeV},'' \href{http://dx.doi.org/10.1016/j.physletb.2014.07.053}{{\em Phys.
  Lett.} {\bf B736} (2014)  371--397},
\href{http://arxiv.org/abs/1405.3886}{{\tt arXiv:1405.3886 [hep-ex]}}.

\bibitem{Khachatryan:2015wza}
{\bf CMS} Collaboration, V.~Khachatryan {\em et al.}, ``{Searches for
  third-generation squark production in fully hadronic final states in
  proton-proton collisions at $ \sqrt{s} = 8$ TeV},''
  \href{http://dx.doi.org/10.1007/JHEP06(2015)116}{{\em JHEP} {\bf 06} (2015)
  116},
\href{http://arxiv.org/abs/1503.08037}{{\tt arXiv:1503.08037 [hep-ex]}}.

\bibitem{Baak:2012kk}
M.~Baak, M.~Goebel, J.~Haller, A.~Hoecker, D.~Kennedy, {\em et al.}, ``{The
  Electroweak Fit of the Standard Model after the Discovery of a New Boson at
  the LHC},'' \href{http://dx.doi.org/10.1140/epjc/s10052-012-2205-9}{{\em
  Eur.Phys.J.} {\bf C72} (2012)  2205},
\href{http://arxiv.org/abs/1209.2716}{{\tt arXiv:1209.2716 [hep-ph]}}.

\bibitem{Baak:2014ora}
{\bf Gfitter Group} Collaboration, M.~Baak {\em et al.}, ``{The global
  electroweak fit at NNLO and prospects for the LHC and ILC},''
  \href{http://dx.doi.org/10.1140/epjc/s10052-014-3046-5}{{\em Eur.Phys.J.}
  {\bf C74} (2014)  3046},
\href{http://arxiv.org/abs/1407.3792}{{\tt arXiv:1407.3792 [hep-ph]}}.

\bibitem{Beringer:1900zz}
{\bf Particle Data Group} Collaboration, J.~Beringer {\em et al.}, ``{Review of
  Particle Physics (RPP)},''
\href{http://dx.doi.org/10.1103/PhysRevD.86.010001}{{\em Phys.Rev.} {\bf D86}
  (2012)  010001}.

\bibitem{Davier:2010nc}
M.~Davier, A.~Hoecker, B.~Malaescu, and Z.~Zhang, ``{Reevaluation of the
  Hadronic Contributions to the Muon g-2 and to alpha(MZ)},''
  \href{http://dx.doi.org/10.1140/epjc/s10052-012-1874-8,
  10.1140/epjc/s10052-010-1515-z}{{\em Eur.Phys.J.} {\bf C71} (2011)  1515},
\href{http://arxiv.org/abs/1010.4180}{{\tt arXiv:1010.4180 [hep-ph]}}.

\bibitem{ATLAS:2014wva}
{ATLAS, CDF, CMS, and D0 Collaborations}, ``{First combination of Tevatron and
  LHC measurements of the top-quark mass},''
\href{http://arxiv.org/abs/1403.4427}{{\tt arXiv:1403.4427 [hep-ex]}}.

\bibitem{Aad:2014aba}
{\bf ATLAS} Collaboration, G.~Aad {\em et al.}, ``{Measurement of the Higgs
  boson mass from the $H\rightarrow \gamma\gamma$ and $H \rightarrow ZZ^{*}
  \rightarrow 4\ell$ channels with the ATLAS detector using 25 fb$^{-1}$ of
  $pp$ collision data},''
  \href{http://dx.doi.org/10.1103/PhysRevD.90.052004}{{\em Phys.Rev.} {\bf D90}
  (2014) no.~5, 052004},
\href{http://arxiv.org/abs/1406.3827}{{\tt arXiv:1406.3827 [hep-ex]}}.

\bibitem{CMS:2014ega}
{\bf CMS} Collaboration, {CMS Collaboration}, ``{Precise determination of the
  mass of the Higgs boson and studies of the compatibility of its couplings
  with the standard model},''
{\em CMS-PAS-HIG-14-009} (2014)  .

\bibitem{Alvarado:2015yna}
C.~Alvarado, A.~Delgado, A.~Martin, and B.~Ostdiek, ``{Dirac Triplet Extension
  of the MSSM},'' \href{http://dx.doi.org/10.1103/PhysRevD.92.035009}{{\em
  Phys. Rev.} {\bf D92} (2015) no.~3, 035009},
\href{http://arxiv.org/abs/1504.03683}{{\tt arXiv:1504.03683 [hep-ph]}}.

\bibitem{Harlander:2002vv}
R.~V. Harlander and W.~B. Kilgore, ``{Production of a pseudoscalar Higgs boson
  at hadron colliders at next-to-next-to leading order},''
  \href{http://dx.doi.org/10.1088/1126-6708/2002/10/017}{{\em JHEP} {\bf 10}
  (2002)  017},
\href{http://arxiv.org/abs/hep-ph/0208096}{{\tt arXiv:hep-ph/0208096
  [hep-ph]}}.

\bibitem{Aad:2016jxj}
{\bf ATLAS} Collaboration, G.~Aad {\em et al.}, ``{Search for new phenomena in
  final states with large jet multiplicities and missing transverse momentum
  with ATLAS using $\sqrt{s} =13$ TeV proton-proton collisions},''
  \href{http://dx.doi.org/10.1016/j.physletb.2016.04.005}{{\em Phys. Lett.}
  {\bf B757} (2016)  334--355},
\href{http://arxiv.org/abs/1602.06194}{{\tt arXiv:1602.06194 [hep-ex]}}.

\bibitem{ATLAS-CONF-2016-010}
``{Search for heavy resonances decaying to a $Z$ boson and a photon in $pp$
  collisions at $\sqrt{s}=13$ TeV with the ATLAS detector},'' Tech. Rep.
  ATLAS-CONF-2016-010, CERN, Geneva, Mar, 2016.
\newblock \url{http://cds.cern.ch/record/2139795}.

\bibitem{CMS-PAS-EXO-16-019}
{\bf CMS Collaboration} Collaboration, ``{Search for high-mass resonances in
  $Z\gamma \rightarrow e^+e^-\gamma/\mu^+\mu^-\gamma$ final states in
  proton-proton collisions at $\sqrt{s}=13~\mathrm{TeV}$},'' Tech. Rep.
  CMS-PAS-EXO-16-019, CERN, Geneva, 2016.
\newblock \url{http://cds.cern.ch/record/2141740}.

\bibitem{Cohen:2015ala}
T.~Cohen, J.~Kearney, and M.~Luty, ``{Natural Supersymmetry without Light
  Higgsinos},'' \href{http://dx.doi.org/10.1103/PhysRevD.91.075004}{{\em Phys.
  Rev.} {\bf D91} (2015)  075004},
\href{http://arxiv.org/abs/1501.01962}{{\tt arXiv:1501.01962 [hep-ph]}}.

\bibitem{Nelson:2015cea}
A.~E. Nelson and T.~S. Roy, ``{New Supersoft Supersymmetry Breaking Operators
  and a Solution to the $\mu$ Problem},''
  \href{http://dx.doi.org/10.1103/PhysRevLett.114.201802}{{\em Phys. Rev.
  Lett.} {\bf 114} (2015)  201802},
\href{http://arxiv.org/abs/1501.03251}{{\tt arXiv:1501.03251 [hep-ph]}}.

\end{thebibliography}\endgroup
\bibliographystyle{utphys}
\end{spacing}
\end{document}